\pdfoutput=1
\documentclass[twocolumn,10pt]{article}
\usepackage[T1]{fontenc}
\usepackage[utf8]{inputenc}
\usepackage[letterpaper,margin=1.7cm]{geometry}
\usepackage{graphicx}
\graphicspath{{figs/}}
\usepackage{dblfloatfix}
\usepackage{booktabs}
\usepackage{tabularx}
\usepackage{amsmath,amssymb}
\usepackage{textcomp}
\usepackage{gensymb}
\usepackage[numbers,sort&compress]{natbib}
\usepackage{caption}
\usepackage{pdfpages}
\usepackage[hidelinks]{hyperref}

\usepackage{titlesec}
\titleformat*{\section}{\raggedright\normalfont\Large\bfseries}
\titleformat*{\subsection}{\raggedright\normalfont\large\bfseries}

\begin{document}

\twocolumn[{%
\begin{center}
{\LARGE\bfseries How good are universal MLIPs for alloys?\\[2pt] A benchmark is only as trustworthy as its controls}\\[0.7em]
{\large Gus L. W. Hart\textsuperscript{1} and Jacob N. Pursley\textsuperscript{2}}\\[0.4em]{\footnotesize \textsuperscript{1}Department of Physics and Astronomy, Brigham Young University, Provo, UT, USA; \textsuperscript{2}Department of Energy's Kansas City National Security Campus, Kansas City, MO, USA}\\[0.3em]{\footnotesize Correspondence: Gus L. W. Hart (gus.hart@byu.edu)}
\end{center}\vspace{0.7em}
\begin{minipage}{\textwidth}\small\noindent\textbf{Abstract.} Universal machine-learned interatomic potentials (uMLIPs) are ranked by their formation-energy error against a DFT database, as if that database were ground truth, but independent DFT databases disagree. We benchmark fourteen uMLIPs on 1.45 million ordered intermetallics against AFLOW, which none trained on, and compare every energy and stability error with that disagreement. For half the structures, AFLOW, Alexandria, and OQMD differ by less than 6--8 meV/atom, but a minority differ by far more, raising the average to 16--30. That disagreement sets a floor on any model's measurable error. Re-referenced, the best uMLIP agrees with AFLOW as closely as Alexandria does, and the next four sit 2--7 meV/atom above the floor. Once broken elemental references are screened, the energy and stability rankings agree almost everywhere. Three apparent model failures shrink or vanish under the right control: poorly correlated shear moduli, thousands of ``bad'' DFT references, and the leader's inflated lead.\end{minipage}\vspace{1.6em}
}]
\section{Introduction}
Universal machine-learned interatomic potentials (uMLIPs) are atomistic models trained to reproduce density-functional theory (DFT) across much of the periodic table. They predict the energies, forces, and relaxed geometries of crystals at a tiny fraction of the cost of DFT. Their speed is reshaping how candidate materials are screened and discovered. Public leaderboards most often rank them by the mean absolute error (MAE) of predicted formation energies against a reference database such as the Materials Project (MP) \cite{jain2013}, and by convex-hull stability classification, the central metric of Matbench Discovery \cite{riebesell2025}.

Rankings are good for spurring progress, but two assumptions deserve a harder look. The first is that a smaller energy MAE means a model that describes the potential-energy surface (PES) more accurately. However, a model can have a lower formation-energy error simply because its training data used the same per-element energy conventions as the reference database. That shared convention is a bookkeeping advantage, not a better PES. A low energy MAE is also an unreliable predictor of which phases a model will actually place on the convex hull \cite{bartel2020,riebesell2025}.

uMLIPs are fitted, black-box models, and we make no claim about what they represent internally. What we can test is the part of a model's behavior that survives a change of elemental reference. At a fixed composition, any per-element shift cancels exactly, so the energy ordering a model predicts among competing structures is a property of its PES and of nothing else.

The second assumption, that the reference database is ground truth, matters more, and it does not hold. Even with a shared exchange-correlation functional, independent high-throughput DFT databases do not agree exactly with one another about formation energies \cite{hegde2023}, because they differ in pseudopotentials, settings, corrections, and elemental references. Benchmarks almost never compare a model's error with the disagreement between two careful DFT calculations of the same quantity.

We address both benchmarking problems by changing the DFT reference, the chemistry, and the metrics. For an independent reference, we use AFLOW data \cite{curtarolo2012}, a high-throughput DFT database that is independent of the data the models were trained on (MPtrj \cite{deng2023}, OMat24 \cite{barrosoluque2024}, and Alexandria \cite{wang2021,schmidt2022}, none of which includes AFLOW). We focus on \textbf{intermetallics}, a chemically rich and under-benchmarked region of materials space where DFT results are relatively accurate to the underlying physics. AFLOW offers systematically enumerated ordered-prototype libraries and precomputed elastic constants (benchmark section).

Two design choices keep the test clean. First, we compare formation energies rather than total energies and lead with properties that depend only weakly on the elemental reference: the ordering of phases on the convex hull, the curvature of the elastic tensor, the geometric drift of a structure as a model relaxes it, and the energy ordering of polymorphs at fixed composition.  Polymorph ordering is exactly reference-free; hull metrics are immune to a self-consistent change of convention but not to a wrong elemental reference (reference-free section). Each metric is still scored against AFLOW, so nothing here is true by construction.

Second, \textit{we compare every energy and stability error with the DFT-vs-DFT noise floor.} The question is how a model's error compares with the disagreement between the DFT calculations themselves. On our set of intermetallics, three independent DFT databases (AFLOW \cite{curtarolo2012}, Alexandria \cite{wang2021}, and the Open Quantum Materials Database (OQMD) \cite{saal2013,kirklin2015}) agree with one another to 6--8 meV/atom on typical structures. The mean disagreement increases to 16--30 (noise-floor section) because of heavy tails of large disagreements. These floors, not zero, are the realistic target for a model.

We test each suspicious result with a \textit{control}, a second version of the measurement that changes one suspected cause (the geometry, the elemental reference, or the Hamiltonian) and holds everything else fixed. Three apparent model failures shrank or vanished under these controls: shear moduli that tracked AFLOW poorly (reference-free section), some fifteen thousand apparent failures, most of which a third database traces to AFLOW where it can settle them (bad-references section), and the leading model's 7 meV/atom lead, which shrinks to 0.4 at a common geometry (leaderboard section).

We report the results that survive these controls, and the controls themselves. Leaderboards for these models are now numerous, and other benchmarks can reuse the controls more readily than any ranking.

The closest existing benchmark is Matbench Discovery \cite{riebesell2025}, which scores the same kind of models on convex-hull stability, a metric we adopt in the reference-free section. It differs in the reference database. Matbench Discovery generates and relaxes its test structures in the Materials Project's own DFT convention and scores them against the MP hull. It therefore measures how faithfully a model reproduces its training database.

Matbench Discovery uses AFLOW only for its prototype labels and never as an energy reference \cite{riebesell2025}. The two benchmarks are complementary. Matbench Discovery is a broad-chemistry stability test within the training convention. Our benchmark tests transfer to an independent reference on metals, compares every energy and stability error with the DFT-vs-DFT floor, and adds elastic, geometric, and reference-convention diagnostics beyond the convex hull.

Other general-purpose platforms such as MLIP Arena \cite{chiang2025}, LAMBench \cite{peng2025}, and CHIPS-FF \cite{wines2025} score many potentials across many tasks, as do broad multi-model assessments \cite{yu2024,huang2025}, while targeted studies probe elasticity \cite{gao2025}, phonons \cite{loew2025}, mechanical softening \cite{deng2025}, and distribution shift \cite{kreiman2026}. None of them scores against a reference the models never trained on or compares errors with a measured DFT-vs-DFT floor.

We estimate how much of each model's error a change of elemental reference could explain by realigning the model to AFLOW's elemental references, a correction that changes no model parameter and that we use as a diagnostic. Once broken elemental references are screened, the stability ranking agrees with the energy ranking almost everywhere, while forces and symmetry retention order the models differently (reference-free section). Because each metric answers a different question, an objective benchmark reports them side by side.

\section{Results}
\subsection{The benchmark: reference, test set, potentials}
\textit{The reference} is AFLOW, a high-throughput DFT database of several million compounds built largely by combinatorially ``decorating'' a fixed library of crystallographic prototypes with elements from across the periodic table \cite{curtarolo2012,calderon2015}. Three features matter here. First, none of the potentials we test was trained on it. Second, its binary (LIB2) and ternary (LIB3) prototype libraries systematically enumerate ordered intermetallic structures. Formation-energy leaderboards, weighted toward oxides and the Materials Project's Inorganic Crystal Structure Database (ICSD) entries, undersample these chemistries. (AFLOW's own ICSD oxides would add little, since 98\% of its 18,943 ICSD entries containing oxygen have a composition already in MP.)

Third, for a small subset, AFLOW reports elastic constants through AEL \cite{toher2017}. Because elastic moduli measure the curvature of the PES rather than its absolute level, they let us compare a model to DFT without choosing any energy reference at all---a probe of $\sim$3,500 compounds that a formation-energy benchmark cannot provide.

\textit{The test set} is $\sim$1.45 million ordered structures. The core is 1.38M binary and ternary intermetallic prototype decorations. The remainder supplies the elemental references, the lanthanide-alloy and extended-chemistry strata, the elastic subset, and a small quaternary set. We exclude the 1,575 elemental entries that AFLOW did not compute with PBE (Methods).

\textit{The potentials} are fourteen uMLIPs (Table 1), chosen to span different ML architectures and training corpora, since the training data determines which DFT a model has effectively learned. They fall into two generations, an earlier one trained on smaller datasets (MPtrj \cite{deng2023}, MatPES \cite{kaplan2025}, or the MatterSim set \cite{yang2024}) and a newer one trained mainly on the much larger OMat24 corpus.\footnote{eSEN is Meta's energy-conserving Smooth Energy Network, whose 30M-parameter OAM checkpoint is also UMA's backbone \cite{fu2025}. eqV2 is the 86M-parameter EquiformerV2 transformer shipped with the OMat24 release \cite{liao2024,barrosoluque2024}.}

We include the earlier MP generation deliberately, alongside the latest OMat models. If an effect moves all fourteen models together, it almost certainly comes from the benchmark. If it moves only the OMat models or only the MP models, it comes from their training data. The main figures show twelve of the fourteen. The two superseded checkpoints (an early MatterSim and SevenNet-0) appear in the closing leaderboard (Discussion) and the elastic comparison (reference-free section).

\begin{table*}[tb]\centering
\footnotesize
\caption{The fourteen universal MLIPs we benchmark, grouped by training corpus. The \textbf{Trained} column sorts the models into two generations of training data. OMat marks the newer generation, trained mainly on OMat24 (118 million structures). Three of these models (SevenNet-MF, eSEN, eqV2) also use MPtrj and subsampled Alexandria, and DPA-3 uses MPtrj among many other sources. MP marks the earlier generation, trained on smaller datasets (MPtrj, MatPES, or the MatterSim set), with subsampled Alexandria added in one case (MACE-MPA). The two labels are therefore only a rough grouping. The Training-data column gives each model's corpus.}
\begin{tabularx}{\textwidth}{l l X X X}
\toprule
Model & Trained & Architecture & Training data & Ref. \\
\midrule
SevenNet-MF & OMat & E(3)-equivariant message-passing (NequIP-class \cite{batzner2022}), multi-fidelity & OMat24 + MPtrj + Alexandria (multi-fidelity) & \cite{kim2024} \\
GRACE-2L & OMat & Graph atomic cluster expansion (2-layer, tensor) & OMat24 & \cite{bochkarev2025} \\
MACE-OMAT & OMat & Equivariant message-passing, ACE-order messages (MACE) & OMat24 & \cite{batatia2022,batatia2025,barrosoluque2024} \\
DPA-3 & OMat & Attention-based message-passing (DeePMD / DPA) & Multi-source, incl. OMat24 and MPtrj; OMat24 head & \cite{zhang2025} \\
UMA & OMat & Equivariant message-passing (eSEN) + Mixture of Linear Experts & OMat24 (multi-domain, with OMol25, OC20, ODAC25 and OMC25) & \cite{wood2025} \\
Orb-v3 & OMat & Graph network (non-equivariant), energy-conserving & OMat24 & \cite{rhodes2025} \\
eSEN & OMat & Equivariant Smooth Energy Network, energy-conserving (30M) & OMat24 pretrain; MPtrj + sAlex fine-tune & \cite{fu2025} \\
eqV2 & OMat & EquiformerV2 equivariant transformer, direct-force (86M) & OMat24 pretrain; MPtrj + sAlex fine-tune & \cite{liao2024,barrosoluque2024} \\
MACE-MPA & MP & Equivariant message-passing, ACE-order messages (MACE) & MPtrj + Alexandria subset & \cite{batatia2022,batatia2023} \\
MatterSim-5M & MP & Graph network (M3GNet-based) § & MatterSim set (MP-consistent PBE+U) & \cite{yang2024} \\
M3GNet & MP & Graph network (3-body interactions) & MatPES-PBE & \cite{chen2022,kaplan2025} \\
CHGNet & MP & Graph network, charge-informed (uses magnetic moments) & MPtrj & \cite{deng2023} \\
SevenNet-0 \textdagger{} & MP & E(3)-equivariant message-passing (NequIP-class) & MPtrj & \cite{park2024} \\
MatterSim-1M \textdagger{} & MP & Graph network (M3GNet-based) § & MatterSim set (MP-consistent PBE+U) & \cite{yang2024} \\
\bottomrule
\end{tabularx}
\vspace{3pt}
\noindent\begin{minipage}{\textwidth}\footnotesize \textdagger{} an earlier checkpoint, superseded by a newer member of the same family that is in the panel; drawn only in the elastic comparison (Fig. 6) and the closing leaderboard (Fig. 12), not among the headline twelve. § the MatterSim paper presents two backbones (an M3GNet-based graph network and a 182M-parameter Graphormer) and does not name the released 1M/5M checkpoints. The 1M/5M models are the M3GNet-based line per the MatterSim distribution. Checkpoint identifiers, package versions, and loaders are listed in Supplementary Note 1.\end{minipage}
\end{table*}
\textit{The protocol} starts every structure from AFLOW's relaxed geometry and relaxes it again with each potential, so that geometry, formation energy, and relaxation drift are all measured on the same footing (ASE FIRE with a FrechetCellFilter that relaxes cell and positions together; Methods) \cite{bitzek2006,larsen2017}. We use the model's own elemental energies to compute formation energy, so every value starts in that model's native convention. Each element's reference is the lowest energy the model assigns to any structure of that element, after a screen that removes elemental relaxations gone unphysical (Methods; bad-references section returns to them).

\subsection{The DFT-vs-DFT noise floor}
A uMLIP cannot aim for zero error. Two independent high-throughput DFT databases do not return the same formation energy for identical structures. Even if they share the exchange--correlation functional (here, the Perdew--Burke--Ernzerhof functional, PBE \cite{perdew1996}), they differ in the rest of the protocol: pseudopotentials, k-point density, smearing, relaxation tolerances, and elemental references. Those choices shift formation energies measurably. A model's agreement with one such database therefore cannot be ranked more finely than two such databases agree with each other. Before scoring any potential, we measure the DFT-vs-DFT disagreement and treat it as the \textit{noise floor}.

We first measure the energy difference for structures that appear in both AFLOW and a second, independent DFT database. Two structures are the same crystal if pymatgen's StructureMatcher matches them and they have the same space group, and we compare every pair of structures with the same reduced formula across the databases to find them (Methods). Table 2 gives the shared counts; 27,427 structures are common to all three databases. Across these shared structures, the median $\lvert\Delta E_f\rvert$ between AFLOW and Alexandria is 5.8 meV/atom, and between AFLOW and OQMD 8.4 meV/atom (Table 2, Fig. 1, Methods).

\begin{table*}[tb]\centering
\footnotesize
\caption{The DFT-vs-DFT noise floor on the same-crystal metallic set (pairs of structures that StructureMatcher matches and that have the same space group, metallic prototypes only; Methods E), measured two ways. The median $\lvert\Delta E_f\rvert$ is the disagreement on the typical structure. The trimmed MAE is the mean $\lvert\Delta E_f\rvert$ after setting aside the pairs that disagree by more than 1 eV/atom, the same trimmed mean we use to score the models (Methods D). The raw-PBE databases agree to 6--8 meV/atom on the typical structure, but heavy tails of large disagreements (the bad-reference class of the bad-references section, present on both sides of every pair) put their trimmed MAE at 16--30 meV/atom. Restricting to metals changes almost nothing for the raw-PBE pairs, since AFLOW is 99.5\% metallic, but matters for the Materials Project, four fifths of whose entries contain a nonmetal. The Materials Project, the one database that adjusts compound energies after the calculation, sits far outside the raw-PBE range on the median (18.0 versus at most 8.4 meV/atom) and at the top of it on the trimmed MAE (30.3 versus 29.9). OMat24 has no row, because its entries are perturbed configurations rather than relaxed structures and provide no formation energies of relaxed prototypes to compare.}
\begin{tabularx}{\textwidth}{X r r X}
\toprule
DFT pair & median $\lvert\Delta E_f\rvert$ (meV/atom) & trimmed MAE (meV/atom) & shared metallic structures \\
\midrule
AFLOW -- Alexandria & 5.8 & 16.4 & $\sim$448,500 \\
AFLOW -- OQMD & 8.4 & 25.8 & $\sim$30,300 \\
OQMD -- Alexandria & 7.8 & 29.9 & $\sim$56,200 \\
AFLOW -- Materials Project & 18.0 & 30.3 & $\sim$6,400 \\
\bottomrule
\end{tabularx}
\end{table*}
All three use the PBE functional, so this 6--8 meV/atom agreement is the reproducibility of a \textit{protocol}, not of the PBE functional itself. Carefully converged codes that run with matched settings agree on elemental equations of state to about 1 meV/atom \cite{lejaeghere2014,lejaeghere2016}, but a formation energy is far more settings-sensitive. It is a \textit{difference} of total energies of chemically different systems, so the errors that cancel within a single element's equation of state do not cancel here.

The three share this floor because they all report \textit{raw}-PBE energies (\textit{raw} meaning no post-hoc adjustment). The one exception is small. OQMD fits the elemental references of four of our metals (Na, Ti, Sn and Hg) to experiment, which touches 20\% of its pairs with AFLOW and raises its median by about 1 meV/atom (Methods). No AFLOW structure behind these comparisons carries a Hubbard +U (Methods). The one +U-heavy subset, the elastic library, is treated in the reference-free section.

As with the models, the size of the floor depends on which average we quote, the median or the mean. The disagreement distributions are heavy-tailed. Most matched pairs agree to a few meV/atom, but a very small population disagrees catastrophically (the bad-reference class of the bad-references section, present on both sides of every pair). We score every model (leaderboard section) with a \textit{trimmed mean}, the average after setting aside the few structures that disagree by more than 1 eV/atom. The large disagreements that remain below that cutoff put the floor's trimmed mean at 16--30 meV/atom, well above the median.

The median and the mean answer different questions, and neither is wrong. Comparing model means with the median floor, however, puts every model about 11 meV/atom farther from the floor than it is. The leader would sit 8 meV/atom above it instead of 3 below (Supplementary Table 8 gives every model). Every distance from the floor that we quote below therefore uses the same trimmed mean for the model and the floor, on the structures on which the floor is measured (Methods).

This floor might still be illusory. The 6--8 meV/atom median could be merely a per-element reference convention, a constant offset a model could absorb for free, rather than genuine disagreement. We test this possibility by fitting a per-element, composition-linear offset between each database pair and asking how much of the structure-to-structure spread it removes. On the metallic set, a per-element offset fit on 70\% of the chemical systems explains under 7\% of the AFLOW--Alexandria disagreement on the other 30\%, and subtracting it does not lower the floor. The pairs that include OQMD are higher (13--14\%), consistent with OQMD's fitted elemental references.

\begin{figure}[!tb]\centering
\includegraphics[width=\linewidth,height=0.9\textheight,keepaspectratio]{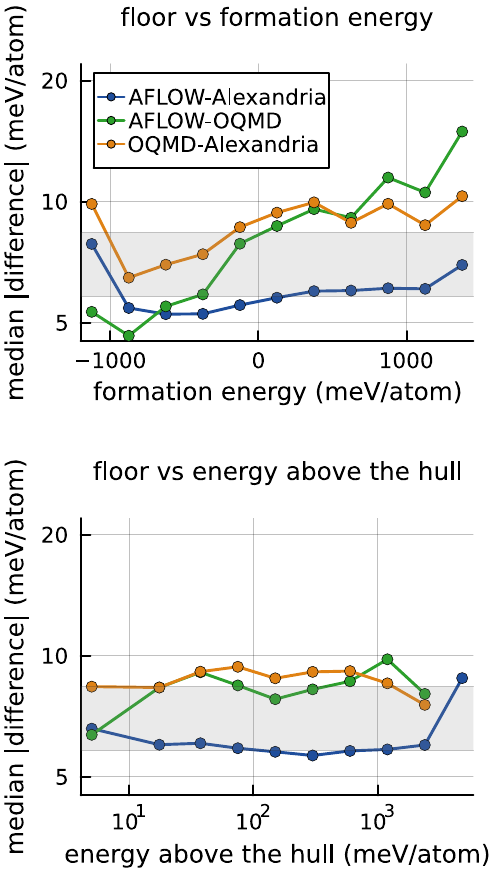}
\caption{The DFT-vs-DFT floor for metallic structures: median $\lvert\Delta E_f\rvert$ over same-crystal pairs for three independent-DFT pairs (AFLOW--Alexandria, 449k, blue; AFLOW--OQMD, 30.4k, green; and OQMD--Alexandria, 27.4k, orange, on the structures common to all three databases, the control in which neither database is AFLOW) versus formation energy (top) and $E_{\mathrm{hull}}$ (bottom). The gray band is the median floor of Table 2. The bins use AFLOW's $E_f$ and $E_{\mathrm{hull}}$, so the OQMD--Alexandria curve holds only the pairs whose crystal AFLOW also holds, and the counts differ from Table 2's. Unlike the model error of Fig. 9, the floors hold a $\sim$5--10 meV/atom band across the whole $E_{\mathrm{hull}}$ range and out to $E_f$ $\approx$ +750 meV/atom (bin-level values in Supplementary Note 5). The reference does not degrade appreciably with distance from the hull, and no single database is the noisy one.}
\end{figure}
The Materials Project, the one database here that adjusts compound energies after the calculation \cite{wang2006,jain2011}, sits farther from AFLOW than the other two. On our intermetallic subset (where none of MP's anion or +U corrections are applied), MP disagrees with AFLOW by 18.0 meV/atom on a typical structure, compared with 6--8 for the raw-PBE databases. A per-element offset removes little of that gap (in-sample R\textsuperscript{2} = 0.06; Supplementary Note 5). The gap is not MP's corrections, which are inactive on metals, and not a re-referenceable convention. It is a genuine protocol difference that we do not isolate further. Hegde et al., in a pairwise comparison of AFLOW, the Materials Project, and OQMD, reached the same conclusion \cite{hegde2023}. Across all chemistries, MP is the common term in the two largest gaps, 105 meV/atom for AFLOW--MP and 87 for MP--OQMD, compared with only 19 for AFLOW--OQMD. Hegde et al. trace these gaps to oxide-dominated references and post-hoc corrections, the very corrections that fall away on metals.

\subsection{The formation-energy leaderboard}
We begin with the familiar metric, each potential's formation-energy error against AFLOW. For each potential we relax every structure, compare its formation energy to AFLOW's, and report the trimmed MAE of the noise-floor section (structures with \textbar{}error\textbar{} > 1 eV/atom are counted separately; Methods). We determine 95\% confidence intervals from a bootstrap that resamples whole chemical systems (Methods D). The rankings and confidence intervals are given in Fig. 2 and Table 3.

The models fall into two tiers, with one model above both. eqV2 leads outright at 18.3 meV/atom, and eSEN is next at 25.5. Behind them the OMat-trained models cluster tightly, 28 to 36. The MP-trained models trail well behind, with MatterSim-5M at 52, M3GNet at 63, and CHGNet at 127. MACE-MPA breaks that split. At 33 it sits in the middle of the OMat-trained models, away from its MP-trained siblings, for a reason we come to below. The separations are far larger than the confidence intervals, none of which is wider than $\pm$2 meV/atom (Table 3).

\begin{figure}[!tb]\centering
\includegraphics[width=\linewidth,height=0.9\textheight,keepaspectratio]{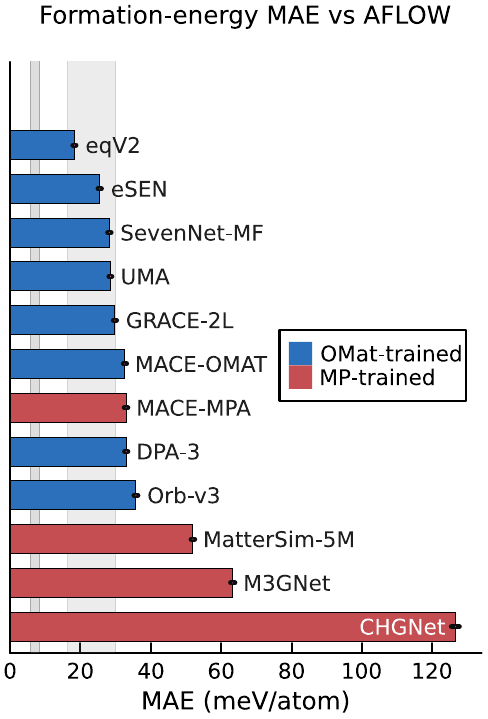}
\caption{Formation-energy MAE vs AFLOW for the twelve headline uMLIPs, sorted best to worst, with 95\% bootstrap confidence intervals. Bars colored by training corpus (blue = OMat-trained, red = MP-trained). The shaded bands mark the median (6--8 meV/atom) and trimmed-MAE (16--30) floors of the noise-floor section; the bars use the same trimmed mean as the wider band. The bars are errors over the whole benchmark, so a bar's distance from a band is only indicative; distances from the floor measured on the structures that define it are in Table 3 (``Above floor'').}
\end{figure}
eqV2's 7 meV/atom lead over eSEN is statistically significant. Before interpreting this as a ranking of PES accuracy, we ask what a low MAE against AFLOW measures. In part, it measures how closely the DFT conventions each model was trained on agree with AFLOW's. eqV2 and eSEN were both fine-tuned on subsampled Alexandria, a database that agrees with AFLOW to within 6 meV/atom on the typical structure (noise-floor section). The fine-tune is worth 1.5--3.6 meV/atom of raw error, almost all of it a per-element offset (Supplementary Note 7).

A second question is how eqV2 reached its score. In the relax-again protocol, it barely moved. Its forces come from a separate \textit{output head}, a final part of the network trained to predict forces directly, rather than from the gradient of its energy. That head treats AFLOW's minima as already converged (Methods). It was therefore scored at or very near the DFT geometry, while the other thirteen models absorbed \textit{relaxation drift}, the change in energy as a structure relaxes away from AFLOW's geometry (Methods). The control that puts every model on the same footing scores each model at AFLOW's geometry without re-relaxing it (single-point), using the energy recorded before its optimizer's first step (Methods).

The single-point control changes two things (Table 3, Single-point). Every model improves, by 4 to 25 meV/atom, and the lead all but vanishes. eqV2's 7 meV/atom margin over eSEN becomes four tenths (14.4 versus 14.8). A paired bootstrap still resolves that order (eqV2 lower by 0.43 meV/atom [0.39--0.48]), but four tenths of a meV is about 3\% of the trimmed floor. The ordering is statistically resolved but too small to matter. Orb-v3 moves the most in rank, from ninth to fifth (35.8 to 20.2). Only CHGNet moves more in meV/atom (126.5 to 101.7). Nearly half of Orb-v3's apparent error (15.6 of 35.8 meV/atom) came from relaxing structures away from AFLOW's geometry rather than from mis-describing the energy at a fixed one.

Two explanations compete for why the MP-trained models trail so far behind. The first is genuine model error, that their potential-energy surfaces are less accurate. The second is bookkeeping. A model trained on Materials Project data inherits MP's per-element energy conventions, its anion and DFT+U corrections, its elemental references. Because a formation energy is computed by subtracting elemental reference energies, any per-element gap between MP's references and AFLOW's shifts every formation energy by a fixed composition-weighted amount. When the model is scored against AFLOW, that inherited convention appears as error even when the underlying PES is sound.

The two explanations make opposite predictions about what happens when we correct the elemental references, so a simple reference correction, which we call \textbf{Rung-0}, can rule bookkeeping out, though it cannot confirm it. We fit one energy constant per chemical element, choosing the constants that minimize the same mean absolute error we score, and subtract them from the model's formation energies. The correction involves no neural network, no weight update, and no per-structure parameter, only one number per element that re-references the model's energies to AFLOW's. The name Rung-0 suggests its place at the bottom of a hierarchy whose higher rungs (per-structure terms, fine-tuning, retraining) would add real fitting capacity. Rung-0 adds nothing beyond a shift of each element's reference energy.

\begin{table*}[tb]\centering
\footnotesize
\caption{Formation-energy MAE against AFLOW (meV/atom): the relaxed protocol, the single-point control, and the Rung-0 re-referencing, ordered by raw MAE.}
\begin{tabular}{l r r r r r}
\toprule
Model & Raw MAE [95\% CI] & Single-point & After Rung-0 & SP + Rung-0 & Above floor \\
\midrule
eqV2 & 18.3 [17.9--18.7] & 14.4 & 16.6 & 12.7 & $-$2.6 [$-$2.8, $-$2.4] \\
eSEN & 25.5 [25.0--26.0] & 14.8 & 23.8 & 13.3 & +2.3 [+1.9, +2.6] \\
SevenNet-MF & 28.2 [27.8--28.7] & 19.4 & 27.0 & 17.9 & +5.8 [+5.5, +6.1] \\
UMA & 28.5 [28.1--29.0] & 17.9 & 25.1 & 13.3 & +4.6 [+4.2, +5.1] \\
GRACE-2L & 29.8 [29.3--30.3] & 21.0 & 28.1 & 20.0 & +7.2 [+6.9, +7.5] \\
MACE-OMAT & 32.7 [32.2--33.1] & 24.2 & 30.7 & 22.2 & +9.8 [+9.4, +10.2] \\
MACE-MPA & 33.0 [32.5--33.5] & 21.6 & 31.3 & 21.2 & +9.0 [+8.6, +9.4] \\
DPA-3 & 33.0 [32.6--33.5] & 23.9 & 30.6 & 22.3 & +10.2 [+9.9, +10.6] \\
Orb-v3 & 35.8 [35.2--36.3] & 20.2 & 33.4 & 17.8 & +11.6 [+11.0, +12.3] \\
MatterSim-5M & 52.0 [51.4--52.5] & 39.4 & 40.7 & 31.6 & +21.3 [+20.7, +22.0] \\
M3GNet & 63.3 [62.7--63.9] & 50.1 & 55.8 & 46.1 & +38.8 [+37.8, +39.8] \\
CHGNet & 126.5 [125.4--127.6] & 101.7 & 72.3 & 60.9 & +54.9 [+53.6, +56.2] \\
\bottomrule
\end{tabular}
\vspace{3pt}
\noindent\begin{minipage}{\textwidth}\footnotesize Units meV/atom. \textbf{Raw MAE [95\% CI]}: trimmed mean over the full metallic set; intervals from a bootstrap over whole chemical systems (Methods). \textbf{Single-point}: the same trimmed MAE at AFLOW's geometry, without re-relaxation (Methods). \textbf{After Rung-0}: corrected MAE on a held-out 30\% of the chemical systems, with one energy constant per element fit on the other 70\% by minimizing the absolute error (Methods F). (Uncorrected, the held-out MAE is within 0.7 meV/atom of the full-set value, so the text's before$\rightarrow$after figures use the full-set raw MAE.) \textbf{Above floor}: after-Rung-0 trimmed MAE minus the 16 meV/atom AFLOW--Alexandria floor of the noise-floor section, both on the held-out structures that define the floor (Methods). Intervals are paired 95\% bootstrap intervals over chemical systems, with the per-element constants held at their fit. A negative entry means the model agrees with AFLOW more closely than Alexandria does on those structures. \textbf{SP + Rung-0}: both controls at once, an estimate of structure-dependent error at the reference geometry (Rung-0 fit and scored on the single-point errors, on the same held-out split as After Rung-0).\end{minipage}
\end{table*}
Two safeguards make Rung-0 a sharp diagnostic rather than a fit (Methods F). The first, a capacity bound, limits what Rung-0 \textit{could} remove to one constant per element, which can shift a model's reference frame but cannot reshape its PES. Rung-0 still removes anything linear in composition: a difference of reference convention, a broken elemental reference, or an element-wise bias in the model. It cannot distinguish them, so its gain is the most that any change of elemental reference could remove, not a measured convention.

The second, the held-out split, shows that what Rung-0 \textit{did} remove is transferable. The $\sim$50 constants are fit on 70\% of the chemical systems and scored on the other 30\% (Methods F; fold robustness in Supplementary Note 7), so they cannot flatter the model by absorbing in-sample noise. The test is therefore one-sided. If Rung-0 leaves a model's error unchanged, that error is not a per-element offset. If Rung-0 removes part of it, that part could be a convention or an element-wise bias.

Re-referencing removes little where a model's error has no per-element part. For nine of the twelve models, which are already nearly aligned with AFLOW's reference frame, it removes only 1.2 to 3.4 meV/atom (Table 3).

Three of these nine are aligned only once the elemental references are screened. UMA, eSEN, and MACE-MPA each give their lowest energy for some element to a structure that is not a sound bulk crystal (bad-references section). That structure becomes the element's reference (benchmark section), so every compound containing the element is mis-scored by the same per-element amount. Rung-0 detects and removes this signature but cannot tell such a mistake from a genuine inherited convention. Discarding such structures before the elemental reference is set (bad-references section; Methods) removes the error rather than absorbing it. The raw MAEs fall, MACE-MPA's from 48.2 to 33.0, eSEN's from 38.6 to 25.5, and UMA's from 35.0 to 28.5. All three then behave like every other aligned model.

The three models trained without Alexandria or OMat24 carry a large per-element component on top of a large structure-dependent error. CHGNet is the clearest case. Re-referencing lowers its error from 127 to 72 meV/atom. Two fifths of its apparent error (43\%) is removable with one constant per element, without changing any weight. Those constants barely track the differences between the Materials Project's and AFLOW's elemental references, the convention of CHGNet's training data (r = 0.19, p = 0.21; Supplementary Note 7). They are therefore more likely an element-wise bias in the model than an inherited convention. The 72 that remains is structure-dependent error that no per-element correction can erase. M3GNet and MatterSim-5M behave the same way, improving to 56 and 41 meV/atom.

\begin{figure}[!tb]\centering
\includegraphics[width=\linewidth,height=0.9\textheight,keepaspectratio]{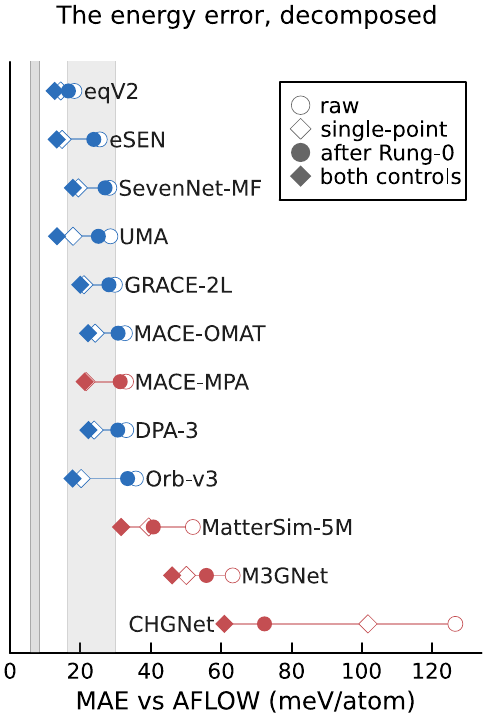}
\caption{The energy error decomposed model by model, sorted by raw MAE (the data of Table 3). Open circle = raw (relaxed); open diamond = single-point at AFLOW's geometry (drift removed); filled circle = after Rung-0 (per-element part removed); filled diamond = both controls. Shaded bands as in Fig. 2.}
\end{figure}
The residual after Rung-0 contains no per-element offset, so it is an estimate of each model's structure-dependent error. Ranked on this residual, the group behind the leader forms a four-way near-tie. eSEN (24), UMA (25), SevenNet-MF (27), and GRACE-2L (28) span 4 meV/atom. Those gaps are smaller than the 6--8 meV/atom uncertainty of the reference itself, so at this scale a ranking measures agreement with AFLOW's particular protocol, not PES accuracy. An equally legitimate raw-PBE reference could plausibly put these four in a different order.

Even after re-referencing, eqV2's lead over eSEN, about 7 meV/atom, sits at the edge of that resolution.

On the structures that define the floor, the re-referenced leader sits 3 meV/atom below the AFLOW--Alexandria floor, eSEN 2 above it, and UMA, SevenNet-MF, and GRACE-2L 5 to 7 above it, all inside the 16--30 band at which the independent raw-PBE databases disagree (Table 3; noise-floor section). These distances depend more on how strictly structures are matched when the floor is built than on sampling. Across the matching tolerances we tried they move by up to about 5 meV/atom for the leading models, compared with intervals of a few tenths, and the leader stays below the floor at every tolerance (by 2.3 to 7.3 meV/atom; Supplementary Table 8).

Before re-referencing, the leader sits just below the floor on the structures that define it, with a trimmed MAE of 15.1 meV/atom compared with the floor's 16.4 and a median of 5.3 compared with 5.8. On the OQMD-matched set its trimmed MAE is 16.9 compared with that pair's 25.8 (Methods). Re-referencing lowers the leader's error on the held-out structures further, from 15.4 to 14.1 meV/atom.

A model can sit below the floor because the floor measures how far two independent DFT databases disagree, not how closely any calculation can match AFLOW. Alexandria and AFLOW differ in pseudopotentials, settings, and convergence, and a potential trained on other PBE data can land closer to AFLOW than Alexandria does. At that point, AFLOW can no longer resolve which of the two is closer to the PBE answer. The floor can be measured only on structures both databases hold, and these structures are in the training data of the models fine-tuned on Alexandria. The OMat24-only checkpoints of eqV2 and eSEN, which never saw Alexandria's labels, also sit below the floor after Rung-0 (Supplementary Note 7), and any memorization of Alexandria compounds is bounded at under 1 meV/atom (generalization section).

We close the error decomposition by applying both controls at once. We fit Rung-0 to the single-point errors, so that neither relaxation drift nor any per-element offset remains (Table 3, SP + Rung-0; the filled diamonds of Fig. 3). Here eqV2's energy margin all but disappears. UMA, whose remaining drift masked it in the relaxed ranking, drops from 17.9 to 13.3 meV/atom, eSEN from 14.8 to 13.3, and eqV2 from 14.4 to 12.7. The fully controlled top three cluster within 0.7 meV/atom. That spread is small next to the floor, so their order says little about which model describes the PES better. Behind them come Orb-v3, SevenNet-MF and GRACE-2L, with MACE-MPA beside them (Table 3).

eqV2 still leads with both controls applied, by about 0.6 meV/atom over eSEN and UMA (Supplementary Table 12). Together, the two controls show that its raw lead came mostly from relaxation drift, which eqV2 barely incurs, rather than from a uniquely better PES at the DFT geometry. The separation that survives most clearly is in hull stability (reference-free section). eqV2 ranks first there both after relaxing and, by a small margin, at AFLOW's geometry.

Rung-0's ingredients are not new. Fitted reference energies are standard practice, from fitted elemental-phase reference energies for every element \cite{stevanovic2012} and OQMD's fitted chemical potentials for 13 elements \cite{kirklin2015} to the Materials Project's per-atom corrections for anions and +U cations \cite{wang2021b}. Machine-learned potentials use them too. MACE shifts each atom's energy by the energy of the isolated atom \cite{batatia2023}, CHGNet normalizes its training energies with elemental reference energies \cite{deng2023}, and multi-fidelity and multi-task training reconcile datasets computed with different settings \cite{kim2025,zhang2024,shiota2024,batatia2025}. Composition-only baselines are textbook \cite{meredig2014,ward2016,jha2018,goodall2020}. A single energy per element is also the lowest-order term of the cluster expansion of alloy configurational energy \cite{connolly1983,sanchez1984,defontaine1994,vandewalle2002,barrosoluque2024b}, the framework on which the formation energies of ordered intermetallics are conventionally built. What is new is using Rung-0 as a \textit{diagnostic}, which tells us what a leaderboard score means without changing the model's PES.

\subsection{Reference-free metrics: independence, not reordering}
\begin{table*}[tb]\centering
\footnotesize
\caption{Stability against AFLOW's hull (stable = within 25 meV/atom), ordered by hull-F1, on the 1,407,872 structures every model scores (Methods G). The first columns use each model's energies after relaxing; the single-point columns use its energies at AFLOW's geometry, the control of the leaderboard section.}
\begin{tabularx}{\textwidth}{X r r r r r r}
\toprule
Model & Precision & Recall & Hull-F1 [95\% CI] & $E_{\mathrm{hull}}$ MAE & Hull-F1, single-point [95\% CI] & $E_{\mathrm{hull}}$ MAE, single-point \\
\midrule
eqV2 & 0.91 & 0.94 & 0.921 [0.910, 0.929] & 18.7 & 0.930 [0.920, 0.938] & 14.7 \\
eSEN & 0.85 & 0.95 & 0.893 [0.882, 0.902] & 26.5 & 0.923 [0.913, 0.932] & 15.6 \\
\textit{Alexandria (second DFT)} & 0.91 & 0.89 & 0.901 & --- & --- & --- \\
UMA & 0.84 & 0.91 & 0.872 [0.860, 0.883] & 27.1 & 0.897 [0.883, 0.909] & 15.7 \\
SevenNet-MF & 0.81 & 0.93 & 0.868 [0.857, 0.879] & 29.0 & 0.901 [0.890, 0.911] & 19.6 \\
GRACE-2L & 0.77 & 0.93 & 0.843 [0.827, 0.856] & 31.4 & 0.874 [0.860, 0.886] & 22.6 \\
MACE-MPA & 0.77 & 0.92 & 0.839 [0.826, 0.851] & 35.5 & 0.878 [0.867, 0.888] & 23.9 \\
Orb-v3 & 0.78 & 0.90 & 0.837 [0.825, 0.849] & 35.8 & 0.875 [0.863, 0.886] & 19.8 \\
MACE-OMAT & 0.75 & 0.90 & 0.821 [0.806, 0.835] & 34.0 & 0.857 [0.843, 0.870] & 25.2 \\
DPA-3 & 0.74 & 0.90 & 0.815 [0.799, 0.830] & 34.0 & 0.847 [0.831, 0.861] & 24.9 \\
MatterSim-5M & 0.62 & 0.91 & 0.736 [0.717, 0.751] & 53.0 & 0.777 [0.759, 0.792] & 41.0 \\
M3GNet & 0.51 & 0.78 & 0.616 [0.594, 0.637] & 62.7 & 0.653 [0.631, 0.674] & 50.8 \\
CHGNet & 0.31 & 0.84 & 0.448 [0.426, 0.469] & 113.1 & 0.522 [0.499, 0.543] & 91.8 \\
\bottomrule
\end{tabularx}
\vspace{3pt}
\noindent\begin{minipage}{\textwidth}\footnotesize Units: $E_{\mathrm{hull}}$ MAE in meV/atom. \textbf{Hull-F1} is the F1 score for classifying a structure as stable (within 25 meV/atom of the hull), with AFLOW's hull as truth and the model's own hull as its prediction; intervals resample chemical systems. \textbf{$E_{\mathrm{hull}}$ MAE} is the mean absolute error in the energy above the hull. Both are blind to the per-element reference convention of the leaderboard section if it is self-consistent, in the sense set out above. The italicized row is the DFT-vs-DFT ceiling. We score Alexandria as a ``model'' against the AFLOW hull, on the shared metallic structures that are the same crystal in both (Methods), and it reaches hull-F1 0.901. Because that row uses a different structure set than the model rows, we also recompute the ceiling and the models on the shared structures (Methods). There the ceiling is 0.913, and eqV2 (0.936) and eSEN (0.923) clear it (Supplementary Note 9). This does not make eqV2 and eSEN more accurate than DFT. A model that agrees with AFLOW more closely than Alexandria does is matching AFLOW's protocol. eqV2 leads eSEN by 0.028 [0.023, 0.033] relaxed and by 0.007 [0.002, 0.011] at AFLOW's geometry (paired differences). Behind eSEN, most adjacent pairs cannot be distinguished, because their intervals overlap.\end{minipage}
\end{table*}
The leaderboard section raises a concern that part of a model's standing on the formation-energy leaderboard is an accident of a reference convention. Rung-0 can remove a per-element offset only after the fact, and only if one chooses to apply it, which leaderboards as ordinarily reported do not. A cleaner route is to lead with properties that need no elemental energy reference, so that bookkeeping cannot affect them. Polymorph ordering is the cleanest such property, and the only energy-based metric that is exactly invariant. Competing polymorphs share a composition, so a per-element shift adds exactly the same constant to every member of the set and cannot change the ordering.

The convex hull is nearly as good. Stability depends on a structure's energy \textit{relative to competing phases of the same elements}, and a per-element shift cancels in that difference as long as the elemental endpoints shift with the compounds. A shift applied to the compounds alone does not cancel. By construction, a model's elemental reference is its own lowest-energy elemental structure, which sits at $E_f = 0$ and anchors that endpoint. If that structure is the right one, the cancellation holds and the metric is blind to the convention. If it is the wrong one, the hull is anchored to a bad structure, and phases move on and off it. Hull metrics are invariant to a self-consistent convention but not immune to a mis-set reference. MACE-MPA's Fe reference before screening was one (bad-references section).

\subsection{Stability on the convex hull}
Matbench Discovery scores a potential by how well it classifies thermodynamic stability \cite{riebesell2025}. We use the same kind of metric, built against AFLOW's hull rather than the Materials Project's, and call the resulting score \textit{hull-F1}.

For every structure we ask whether it is stable. AFLOW's energies give the true hull, and each model's own energies give its predicted hull. A classification can err in two ways, and \textit{precision} and \textit{recall} measure them. Precision is the fraction of the structures a model calls stable that truly are, and recall is the fraction of the truly stable structures it finds.\footnote{One could instead weight recall more heavily (an F2 score) if missing a stable phase were costlier than a false alarm. Here we treat both errors as equally costly, so we use the balanced F1.} The F1 score is their harmonic mean, so it is high only when both are high.

We call a structure stable if it lies within 25 meV/atom of the hull, a few times the 6--8 meV/atom DFT-vs-DFT floor (noise-floor section), so no model is asked to resolve stability more finely than two DFT databases agree. The choice of window does not drive the result, since eqV2 leads at every window from 10 to 50 meV/atom (Supplementary Note 9).

Broken elemental references make the energy and stability rankings look more different than they are. If we rerank the twelve models by stability instead of by energy, a model moves two places on average with the unscreened references, and less than one with them screened. These averages score each model on all the structures it relaxed, not on the common set of Table 4.

eqV2 is first in both rankings and eSEN second. Both gaps are resolved, 7 meV/atom on energy and 0.028 [0.023, 0.033] on hull-F1 (0.921 versus 0.893). At AFLOW's geometry, the single-point control of the leaderboard section, the hull-F1 gap narrows to 0.007 [0.002, 0.011] (0.930 versus 0.923) but stays resolved. Only two models move by more than one place. Orb-v3 ranks ninth on energy but seventh on stability, and MACE-OMAT moves the other way, from sixth to eighth.

The stability metric is useful because a self-consistent change of elemental reference can move the energy ranking but not the stability ranking. A broken reference can move the stability ranking as well (above). Once the references are screened, the two rankings agree almost everywhere.

The divide between the OMat and MP models (leaderboard section) survives the change of metric and is wider here than on energy. The OMat-trained models and MACE-MPA score hull-F1 0.82--0.92, and the other MP-trained models 0.45--0.74. CHGNet is the extreme case. Its recall of 0.84 means it finds most stable phases, but its precision is only 0.31, so more than two of every three structures it calls stable are not. Precision matters for screening new materials, where the hull is the filter. A model this imprecise suggests many candidates that are actually unstable.

\subsection{Residual forces}
Residual forces need no elemental reference. At AFLOW's relaxed geometry the DFT forces are essentially zero, so the force a model reports there measures how well its PES agrees with DFT's near the minimum. On this metric the panel does not split by training set. Ten of the twelve models report a median maximum residual force of 0.052 eV/\AA{} or less. Eight of them lie between 0.040 and 0.052 (UMA lowest of those), and eSEN is lower still at 0.022.

eqV2 reports a median below 0.001 eV/\AA{}. As the leaderboard section noted, it predicts forces with a separate output head, so a near-zero force at a DFT minimum partly reflects what that head was trained to output. (On the energy ranking, the single-point control of the leaderboard section corrects for the same asymmetry.)

Only M3GNet and CHGNet stand apart, near 0.10 eV/\AA{}, with about half of their structures above 0.1 eV/\AA{} compared with a quarter to a third for the other models. The two models the hull metric ranked last are also the two whose PES is farthest from DFT's at the minimum.

\subsection{Space-group retention}
After relaxing each structure, we ask whether it keeps its space group. This is a check on \textit{geometry}, separate from the energy-based hull-F1. At first sight, space-group retention separates the models sharply. Most of that separation, however, comes from the symmetry tolerance (\texttt{symprec}) used to assign the space group. As we loosen \texttt{symprec} (Fig. 4), the models converge. Over the whole benchmark, and leaving eqV2 aside, retention at a tight 0.01 \AA{} runs from 77.5\% (UMA) to 92.0\% (MatterSim-5M). At the conventional 0.1 \AA{} it runs from 88.6\% to 92.2\% (all fourteen models at every tolerance in Supplementary Note 10). eqV2 retains 99\% at every tolerance because it barely relaxes (leaderboard section).

UMA, eSEN, and Orb-v3 lose the most at the tight tolerance (77.5--84.0\%) and gain the most as it loosens, the signature of small distortions that a looser tolerance forgives. A small real difference remains at 0.1 \AA{}, where every interval is about $\pm$0.1 point. Orb-v3 retains the fewest of the OMat-trained models (90.3\%), and the four oldest MP-trained models the fewest overall (88.6--89.0\%). But the large apparent spread came from one analysis choice, so we report retention as a function of tolerance, never as a single number.

\begin{figure}[!tb]\centering
\includegraphics[width=\linewidth,height=0.9\textheight,keepaspectratio]{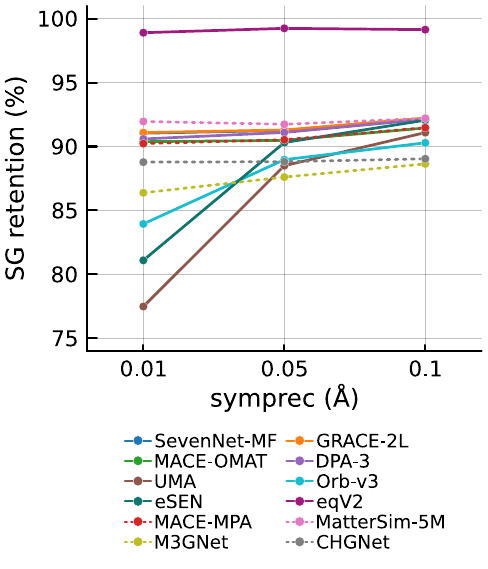}
\caption{Space-group retention versus the spglib symmetry tolerance (symprec), over the whole test set (all 1,456,033 entries, including 1,575 non-PBE elemental entries that the energy metrics drop; intervals of about $\pm$0.1 point are smaller than the markers). A structure is retained when its relaxed space group equals AFLOW's, both assigned at the same tolerance. Leaving aside eqV2, which barely relaxes, the spread across models drops from 14.5 points at 0.01 \AA{} to 3.6 at 0.1 \AA{}. Most of it is a \texttt{symprec} artifact, not a fidelity difference.}
\end{figure}
\subsection{Polymorph ordering}
Polymorph ordering asks whether a model ranks the competing structures at one composition in the same energy order as DFT. We score it by the Spearman correlation $\rho$ across each composition's polymorphs and keep the compositions whose AFLOW energies span at least 25 meV/atom. The two-tier divide appears again. The OMat-trained models reproduce AFLOW's ordering at median $\rho$ 0.92--0.98 (eqV2 highest, then eSEN, UMA, SevenNet-MF), MACE-MPA sits with them at 0.92, and the MP generation trails at 0.79--0.88 (CHGNet lowest). At AFLOW's geometry every model orders the polymorphs better, and the leaders converge. eqV2 and eSEN then tie (mean $\rho$ 0.962 each; each finds the lowest polymorph 62\% of the time), so eqV2's lead over eSEN in the relaxed protocol is a relaxation effect.

A stricter test lowers every score. Asked to pick the single lowest-energy polymorph, the best model (eqV2) is right only 59\% of the time, and CHGNet 28\%, counting a tie for lowest as a hit. Both are far better than chance. Random guessing among the polymorphs of each set, whose median size is 25, would be right 8\% of the time. The models capture the \textit{shape} of the energy landscape but often miss its exact minimum.

Alexandria can be scored only on the compositions it shares with AFLOW, which have far fewer polymorphs (10 on average, compared with 21 on the full set). There it finds AFLOW's minimum 91\% of the time. Scored against AFLOW on the same compositions, eqV2 does slightly better (92\%, mean $\rho$ 0.963 versus Alexandria's 0.956), again matching AFLOW's protocol more closely than a second DFT database does, as in the leaderboard section. eSEN is 2 points below Alexandria and the other leading models 4 to 7 points below (Supplementary Note 11).

The pairs that include OQMD agree less well. OQMD and Alexandria share at least three polymorphs in 2,565 compositions and agree at mean $\rho$ 0.85, finding the same minimum 84\% of the time; AFLOW and OQMD, in 2,318 compositions, reach 0.83 and 85\%. Chance is 28--29\%. AFLOW and Alexandria, reweighted to the same set sizes, give 0.95 and 94\%. (We quote means because nearly all of these sets have only three to five members.) On the AFLOW--OQMD compositions, every OMat-trained model and MACE-MPA orders the polymorphs more consistently with AFLOW than OQMD does (mean $\rho$ 0.84--0.93, versus 0.83), and the MP-only models less so.

\begin{figure}[!tb]\centering
\includegraphics[width=\linewidth,height=0.9\textheight,keepaspectratio]{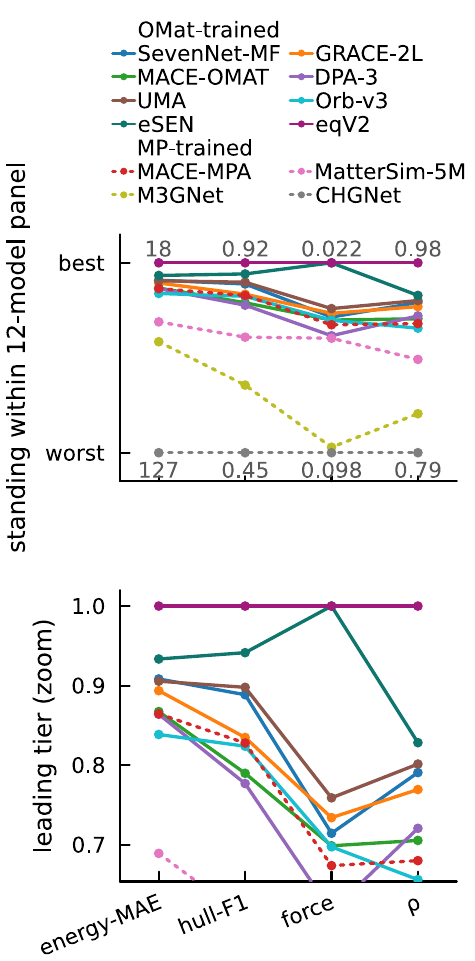}
\caption{The ranking on energy MAE and on three reference-free metrics. Vertical position reflects performance (not merely rank). Each axis is scaled to the panel's best and worst on that metric (gray numbers). The force axis is scaled without eqV2, whose direct-force head reports about 0.0004 eV/\AA{}; eqV2 is drawn at the top. Color identifies the model. Solid lines mark the OMat-trained models and dotted lines the MP-trained ones, and the legend groups them by family. Top, all twelve headline models; bottom, a zoom on the leading tier. Lines cross where models change order, and there are few crossings. eqV2 tops every axis, though its force standing partly reflects its direct-force head (reference-free section). Below eqV2 and eSEN the orders largely agree. MACE-MPA (dotted) sits with the OMat tier on all four.}
\end{figure}
\subsection{Where the rankings differ}
Figure 5 collects these rankings into one picture and spaces the models by how far apart they are, not by rank. Three things stand out. The first and coarsest is the two-tier split above. On the energy axis the gap is very large, with CHGNet at 127 meV/atom compared with the leader's 18.3.

Second, there is little reshuffling inside the leading tier. eqV2 and eSEN rank first and second on all four metrics, and UMA and GRACE-2L hold their places to within one. Orb-v3 is ninth on energy MAE, seventh on the hull and on residual force, and ninth on polymorph ordering. Its relaxation drift (nearly half of its formation-energy error; leaderboard section) affects the reference-free metrics too. At AFLOW's geometry it rises to sixth on the hull and fourth on polymorph ordering. Rung-0 reduces its energy error only slightly (36$\rightarrow$33). The order in the middle of the energy ranking also depends on the cutoff of the trimmed mean, while the leader and the tiers do not. Orb-v3 moves three places between cutoffs of 0.5 and 1.0 eV/atom (Supplementary Note 6; Supplementary Fig. 1).

Third, MACE-MPA does not stay with its group. It is the one dotted curve in the leading tier. We class it with the MP generation, but on the reference-free metrics it rises into the OMat-trained tier.

The clearest comparison for MACE-MPA is with MACE-OMAT, which has the same architecture and differs only in training data \cite{batatia2023,barrosoluque2024}. The two are close on every metric we have, with no consistent winner: hull-F1 0.839 versus 0.821, residual force 0.047 versus 0.045 eV/\AA{}, polymorph ordering $\rho$ 0.920 versus 0.925, and, once the elemental references are screened, energy MAE 33.0 versus 32.7 meV/atom. Before screening, their energy MAEs differ by 16 meV/atom in MACE-OMAT's favor, which would suggest that the larger OMat corpus separates them. In fact, that gap comes from MACE-MPA's broken Fe reference (bad-references section).

MACE-MPA most likely keeps pace with the OMat-trained models because of its training set. It is the one MP-generation model in our panel trained on subsampled Alexandria as well as MPtrj, and Alexandria reports formation energies in the same raw-PBE convention as AFLOW. Capability therefore follows the OMat-versus-MP label only roughly.

\subsection{Elastic response}
The elastic response is another reference-free test. For about 3,500 of our structures, AFLOW's AEL library stores the full DFT elastic tensor \cite{toher2014,toher2017}, from which we read a bulk modulus \textit{K} and a shear modulus \textit{G}. We relax each structure with a model and recompute its tensor by the same finite-strain protocol, which yields a model \textit{K} and \textit{G} for the same compound. Because moduli are curvatures of the PES rather than absolute energies, the comparison needs no reference at all.

Across these compounds, the bulk modulus tracks AFLOW well for every model (log-space correlation \textit{r} from 0.91 to 0.99). The shear modulus does not. Its log-space \textit{r} runs from 0.76 for CHGNet and M3GNet to 0.94 for UMA. Taken at face value, several models predict shear stiffness poorly. Systematic softening has been reported across uMLIPs \cite{deng2025}, and elastic response is now a benchmark target in its own right \cite{gao2025}.

Before blaming the models, we ask what the poorly predicted structures share. It is the \textit{reference Hamiltonian}. AFLOW computes its alloy libraries (LIB2 and LIB3, 95\% of our benchmark) without +U, but elsewhere it applies a Hubbard +U correction \cite{dudarev1998,calderon2015} to compounds containing any of 28 transition and post-transition metals, intermetallics included (Methods H). +U compounds are therefore rare in the benchmark (1.9\%, and none of the structures behind its energy metrics) but make up 78\% of the AEL comparison set. The Materials Project applies +U only to oxides and fluorides of eight of those elements, so for most of these compounds the models were trained on a different Hamiltonian than AFLOW used. The shear response, which measures the response to a change of shape, is more sensitive to that choice than the bulk response to uniform compression.

When we restrict the comparison to the AFLOW structures computed \textit{without} +U (about 795 of them), the shear correlation recovers (Fig. 6), to 0.85--0.98 (CHGNet 0.76 $\rightarrow$ 0.88, MACE-MPA 0.92 $\rightarrow$ 0.98, UMA 0.94 $\rightarrow$ 0.98). Splitting the +U structures by chemistry (Supplementary Note 12) shows shear agreement worst for the +U intermetallics, a treatment the Materials Project never uses, and lowest there for five of the eight models. The picture is not purely one of convention, though. The +U oxides and fluorides of elements the Materials Project treats without +U agree well for most models, while the 89 structures where both databases apply +U are poorly predicted by the older models.

Working in log space shows what kind of error is left (Supplementary Note 12). On the subset without +U, every model's log--log slope against AFLOW's shear moduli lies between 0.91 and 1.02. The older models therefore do not drift further from AFLOW as the modulus grows. They are off by a constant \textit{factor}. CHGNet predicts shear at 0.48$\times$ AFLOW across the board, M3GNet at 0.71$\times$, and SevenNet-0 at 0.73$\times$, while MACE-MPA, GRACE-2L, Orb-v3, and UMA sit at 0.97--1.03$\times$.

A linear correlation would tell a different story for MACE-MPA, \textit{r} = 0.16 on all structures. That value is set almost entirely by one structure, NiB\textsubscript{1}\textsubscript{2}, predicted with a shear modulus of $-$12,719 GPa; without it the linear \textit{r} is 0.66. Its median shear ratio to AFLOW is 0.97.

\begin{figure}[!tb]\centering
\includegraphics[width=\linewidth,height=0.9\textheight,keepaspectratio]{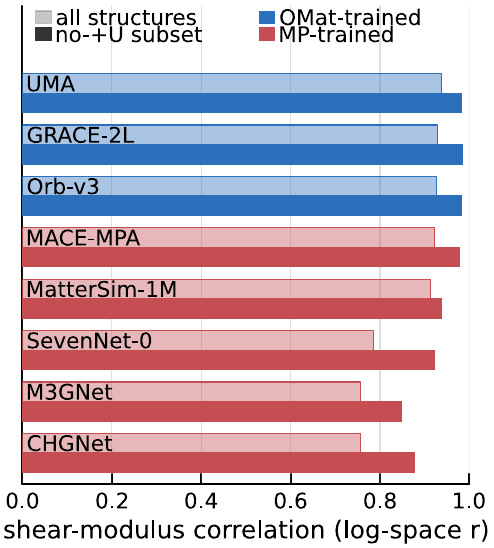}
\caption{Log-space correlation of predicted shear modulus with AFLOW, on all structures (faded) and on the structures AFLOW computed without +U (solid). Structures a model predicts with a non-positive shear modulus are left out (Methods H).}
\end{figure}
Much of the apparent scatter in shear therefore comes from comparing models trained on one Hamiltonian with a reference computed with another. The older models' constant-factor softening (0.48--0.73$\times$) survives the control. A difference in Hamiltonian can look like a model error, so the structures should be stratified by how the reference was computed before an error is blamed on a model's PES. The argument has a limit. Removing the +U structures from the elastic set also removes every compound containing any of the 28 elements AFLOW treats with +U, so we cannot separate +U from element identity or from correlated-electron effects.

The elastic set is not metals-only, since AFLOW's elastic data live mainly on ICSD compounds with nonmetals; a metals-only recut leaves the conclusion intact at lower statistical power (Supplementary Note 12).

\subsection{``Bad'' references: mostly the benchmark's, sometimes the model's}
\begin{table*}[tb]\centering
\footnotesize
\caption{The controls we apply. Each isolates one way a model-versus-reference comparison can mislead, and several reclassify an apparent model failure as a measured reference or protocol artifact.}
\begin{tabularx}{\textwidth}{X X l X}
\toprule
Control & What it guards against & Section & What it revealed \\
\midrule
Independent reference (AFLOW) & Memorization posing as accuracy & Benchmark & 52--98\% of the alloys are novel to a model, depending on its training corpus, so the benchmark measures transfer at least as much as recall \\
DFT-vs-DFT noise floor & Scoring error against zero & Noise floor & Independent raw-PBE databases agree to only 6--8 meV/atom on the typical structure; below that scale a ranking is not transferable \\
Floor and models under the same trimmed mean, on the same structures & Comparing a median floor with mean model errors & Noise floor & Under the trimmed mean the floor is 16--30 meV/atom, not 6--8; after re-referencing, the leader sits 3 below it, eSEN 2 above, and UMA, SevenNet-MF, and GRACE-2L 5--7 above \\
Clean metallic cut (no oxides or nitrides) & A chemistry confound, anion and +U corrections posing as a metallic convention & Noise floor & MP sits 18.0 meV/atom from AFLOW on a typical structure (the raw-PBE databases 6--8), and a per-element offset removes little of it (in-sample R\textsuperscript{2} = 0.06) \\
System-resampled bootstrap & Pseudo-replication, near-duplicate decorations posing as independent evidence & Leaderboard & Confidence intervals no wider than $\pm$2 meV/atom \\
Single-point scoring at the reference geometry (all fourteen models) & A protocol asymmetry: the direct-force leader barely relaxed, while thirteen models absorbed relaxation drift & Leaderboard & Every model improves by 4--25 meV/atom; the leader's margin over eSEN shrinks from 7 to 0.4 meV/atom \\
Rung-0 per-element re-referencing (fit on 70\% of the chemical systems, scored on the other 30\%) & Error linear in composition (a reference convention, a broken reference, or element-wise bias) counted as structure-dependent error & Leaderboard & 1--3 meV/atom for nine of the twelve models; 7--54 meV/atom for the three trained without Alexandria or OMat24 (CHGNet 127$\rightarrow$72); the group behind the leader spans 4 meV/atom \\
Symprec tolerance sweep & An analysis parameter posing as a fidelity difference & Reference-free & The spread in retention across models (eqV2 aside) shrinks from 14.5 points to 3.6 \\
Reference-Hamiltonian stratification (drop +U structures) & A Hamiltonian mismatch posing as a model failure & Reference-free & Shear agreement recovers for every model (log-space r 0.85--0.98 without +U; MACE-MPA 0.92 $\rightarrow$ 0.98) \\
Elemental-reference screen (S0 isolated atoms, S1 low outlier, S2 volume change) & A model's own pathological elemental structure setting the reference for every compound containing that element & Benchmark & Ten references across five models; removing them lowers raw MAE by 15 meV/atom for MACE-MPA, 13 for eSEN, and 7 for UMA \\
Consensus test (the models agree and AFLOW stands apart) & Blaming the model when the reference is the outlier & Bad references & Finds 15,278 discrepancies (structures where AFLOW sits far from a tight model consensus) among 1.42M structures, 10,861 after the per-element offset; the count changes smoothly with the threshold \\
Third-database check (Alexandria, OQMD) & An unresolved model-versus-reference disagreement & Bad references & Where Alexandria has the same structure (3,794 discrepancies), it sides with the models in 83\% of cases (80\% when the whole test is rebuilt on the six MP-generation models); 643 are genuine model errors, concentrated in 4d/5d refractory metals \\
Cross-database validation (OQMD-only compounds) & A conclusion that is AFLOW-specific & Generalization & The same leading models on the 12,409 OQMD compounds absent from AFLOW whose elements carry unary references (4,821 also absent from every training set), so the lead reflects the models and not AFLOW \\
\bottomrule
\end{tabularx}
\end{table*}
Most disagreements between a model and AFLOW are a few tens of meV/atom or less. A small population of structures miss by whole electron-volts. There, either the model or the reference is badly wrong, and deciding which needs evidence from outside the pair.

We therefore look for the structures where AFLOW disagrees with the \textit{consensus} of eight of the models rather than those a single model gets wrong. When the potentials cluster within a few tens of meV and AFLOW sits an electron-volt away, the reference is more likely to be wrong. A structure on which the models agree closely and AFLOW sits far from all of them is a \textit{discrepancy} (the \textit{consensus test}; thresholds in Methods I). The consensus test finds 15,278 discrepancies among the 1.42M structures it can score, and the count changes smoothly with the threshold.

Fifteen thousand bad references would be a serious problem for AFLOW, but part of each disagreement is the per-element reference offset of the noise-floor and leaderboard sections \cite{stevanovic2012}. We fit that offset to the gap between the consensus and AFLOW, subtract it, and apply the same cut again. The discrepancies fall from 15,278 to 10,861. Of the 4,419 structures that drop out, 91\% contain a magnetic 3d element (V, Cr, Mn, Fe, Co, or Ni), compared with 35\% of all structures. These elements carry the largest fitted offsets (Mn $-$119, Cr $-$103, and Fe $-$87 meV/atom). Magnetism itself does not make the models worse. No model has a larger trimmed MAE on magnetic structures than on nonmagnetic ones (Supplementary Note 16).

The offset alone does not explain the 4,419 discrepancies that drop out. After we remove the offset, each of them still sits more than 200 meV/atom from the consensus. Each drops out because the offset carries it just below the 0.3 eV/atom cut. Where Alexandria has the same structure, it sides with the models on 86\% of the dropouts, so most of them are AFLOW disagreements a little below the threshold. None of the discrepant structures, before or after re-referencing, is computed with +U (the alloy libraries carry none; Methods), so none of these discrepancies can be the Hamiltonian mismatch of the reference-free section.

The 10,861 discrepancies that remain after re-referencing are real disagreements, and for these we consult a third database. Where a discrepant structure exists in another DFT database, we ask whether that database agrees with the models or with AFLOW. Alexandria is the most useful third database, because it shares AFLOW's raw-PBE convention (noise-floor section). We count a result only when the other database holds the same crystal as the discrepant AFLOW structure (Methods). Alexandria has the same structure for 3,794 of the 10,861. It sides with the models on 3,151 of them (83\%), with a median disagreement of 13 meV/atom with the model consensus compared with 412 meV/atom with AFLOW. When a second independent DFT database agrees with the models against AFLOW, the reference is almost certainly the one at fault.

The result depends little on which models form the consensus. We rebuild the whole test from the six MP-generation models alone, so that no model in it was trained on OMat24 (one of them, MACE-MPA, saw a subset of Alexandria). Alexandria then has the same structure for 3,794 discrepancies and sides with the models on 3,032 of them, 80\% (Methods). OQMD has the same structure for 322 of the 10,861 discrepancies that remain after re-referencing, and sides with the models on 301 of them (93\%). The Materials Project has the same structure for 97 of the remaining 10,861 and sides with the models on 96, but about half of these are elemental crystals. Where a third database has the same structure, it sides with the models about five times in six. Of the remaining 10,861, 7,067 have no same-crystal Alexandria entry, and for those we cannot say whose error it is.

We also ask whether Alexandria has bad references of its own. We apply the same consensus test to the 448,397 same-crystal AFLOW--Alexandria metallic pairs whose AFLOW structure the models scored, now measuring Alexandria's distance from the models. The models were run on the AFLOW structure, and every pair is the same crystal, so the models' energies apply to Alexandria's entry as well. In 435 pairs the test finds a discrepancy for Alexandria and none for AFLOW, and StructureMatcher confirms the match again on the structure behind Alexandria's energy for 426 of them. We count these 426 as genuine Alexandria errors (Supplementary Note 13).

These 426 are nearly all of one kind. In all but two of them, Alexandria's energy is too high (by a median of 426 meV/atom). In 362, AFLOW finds a magnetic ground state where Alexandria's entry carries no moment. (Supplementary Table 18 gives the chemistry, sources and third-database checks.) On the same pairs and with the same structure test, AFLOW is the outlier in 2,768 pairs, so Alexandria has fewer bad references than AFLOW, and most of Alexandria's are a missed magnetic state.

In 643 of the 3,794 cases Alexandria settles, the two DFT databases agree and the models are wrong. These are genuine model errors. The models share these errors, because a discrepancy requires a tight consensus among them (Methods I) that sits more than 0.3 eV/atom from AFLOW. The errors concentrate in the 4d/5d refractory metals. Of the 643, 437 (68\%) contain a 4d/5d refractory element, compared with 874 (28\%) of the 3,151 on which the models are right. Platinum appears in 139 of the 643 and in 182 of the 3,151, iridium in 79 and 127, and several other refractory metals are over-represented as well (Supplementary Note 13). Manganese and chromium are comparatively rare among the errors (13\% and 11\% of the 643, compared with 45\% and 29\% of the 3,151), so this is not a magnetic-3d effect.

One class of bad reference belongs to the models. The third-database check above cannot see this class, because it asks only whether a model is right about a \textit{compound's} energy, not about the elemental reference the model subtracted to get that energy. MACE-MPA places a diamond-Fe cell 0.94 eV/atom below every sound Fe structure (the cell also collapses 30\% in volume). DPA-3 places a bcc Gd cell more than an electron-volt below its own hcp Gd (the cell expands 59\%), and eSEN places a Cs cell about 3.6 eV/atom below its sound Cs structures (the cell expands 44\%). UMA (Pd) and eSEN (Ti, Sn, V, Ga) also score isolated atoms from AFLOW's unary set below the bulk crystal. Each model puts a minimum where none belongs, and because the elemental reference is the lowest energy a model assigns to that element, one such structure mis-references every compound containing that element.

No third database can settle these cases, because AFLOW is right, and so are the model's compound energies. Only its elemental reference is wrong. We find these structures with the elemental-reference screen (Methods). The leaderboard section shows what removing them does to the raw error of MACE-MPA, eSEN, and UMA.

Three controls thus take 15,278 apparent model failures to 643 confirmed model errors, and leave 7,067 unsettled.

Excluding the 10,861 discrepancies that remain after re-referencing, or all 15,278, lowers every model's trimmed MAE by 2.5--4.0 meV/atom. On the structures that define the floor, the models and the floor drop by similar amounts (1.8--2.9 and 1.8--2.4 meV/atom), so no ranking and no comparison with the floor changes (Supplementary Note 14).

\subsection{Generalization}
A model trained on a very large dataset might top the leaderboard because it generalizes, or only because it has already seen structures in the test set. How these potentials behave under such distribution shift is under active study \cite{kreiman2026}. AFLOW's prototype libraries let us tell the two cases apart. A structure is \textit{familiar} to a model if it appears in that model's training corpus, and \textit{novel} if it does not. Detecting familiarity across a large database is the hard part. We use the same test as for the noise floor: a structure is present if StructureMatcher matches it to a training structure with the same space group. We compare every benchmark structure with every training structure of the same reduced formula, so no match is missed for want of a lookup key (Methods J).

Familiarity is a property of each model, so we tag each structure separately for each model, against public corpora that stand in for that model's training data (Methods J). The familiar fraction separates cleanly by training corpus. Of our alloys, 48.1\% are familiar to the four models trained on the Materials Project together with Alexandria (MACE-MPA, eqV2, eSEN, SevenNet-MF), 39.6\% to DPA-3, trained on OMat24 and the Materials Project, 38.9\% to the four trained on OMat24 alone, and \textbf{2.3\%} to the five trained on the Materials Project only. For OMat24, whose training rows are perturbed configurations, we match against the Alexandria structure each row was generated from. Of the 1,419,259 scored benchmark structures, 552,041 match one, and every one of those also matches an Alexandria structure (Methods). A single pooled figure would be correct for the first four models and wrong for the other ten, by up to 46 percentage points. About half or more of the benchmark is novel for every model (52\% for the four trained on the Materials Project and Alexandria, 60\% for DPA-3, 61\% for the four trained on OMat24 alone), and nearly all of it (98\%) for the MP-trained ones.

To measure how much accuracy a model loses on a structure it has never seen, we compare the familiar structures with every novel one, whose structure is not in the model's training data. Most novel structures have a composition that is in training, with a different structure. Fewer have a new composition in a chemical system that is in training, and for the models trained without OMat24 or Alexandria many sit in chemical systems absent from training altogether (Methods J). For the OMat-trained models, the mean error is 16--29 meV/atom on familiar structures and 5.3 to 11.6 meV/atom higher on novel ones (Fig. 7). eqV2 loses the least (+5.3) and eSEN the most (+11.6), and the MP-trained MACE-MPA loses 15.6. The gap is largest, 12.3 to 22.1 meV/atom, on the novel structures with a new composition in a familiar chemical system, 8--12\% of each OMat-trained model's novel set, because those sit far above the hull (below).

The familiar-structure error of $\sim$24 meV/atom sits inside the trimmed-mean band at which independent DFT databases agree (16--30, noise-floor section). A model that exactly reproduced the DFT it was trained on would score the same. The familiar structures alone therefore cannot separate memorization from skill, but the novel ones can, because memorization would show up as a large jump in error on novel structures. The rise we see is real but moderate, a quarter to three fifths of the familiar-structure error, and far smaller than that of the models trained without OMat24 or Alexandria.

The MP-trained models lose more, CHGNet most at +78, but they were the least accurate to begin with. Memorization does not explain what separates the OMat-trained models from the MP-trained ones, since the largest-data models have the lowest error of the panel on novel structures.

\begin{figure}[!tb]\centering
\includegraphics[width=\linewidth,height=0.9\textheight,keepaspectratio]{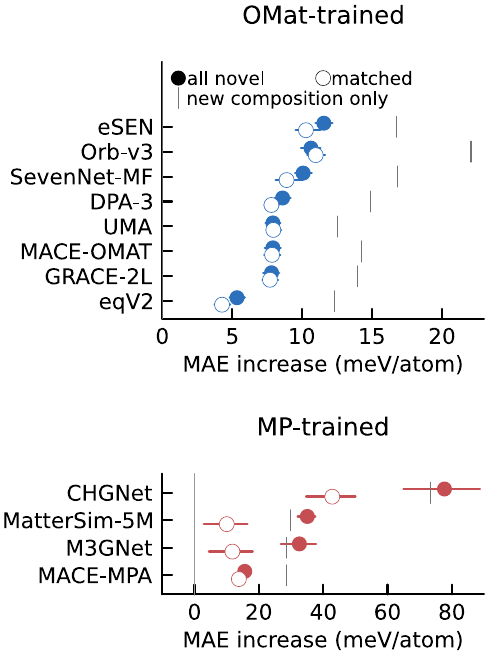}
\caption{Generalization gap: the increase in MAE from familiar structures (the same crystal is in the model's training data) to novel ones. Filled dots, all novel structures; open dots, the same gap matched on hull distance and number of elements; gray ticks, the novel structures with a new composition in a familiar chemical system alone. Whiskers are 95\% intervals from resampling chemical systems. The two panels have different scales. Familiarity is tagged per model, against public corpora that stand in for that model's training data (Methods J).}
\end{figure}
Changing the reference database also tests generalization. We ran all fourteen models on 34,681 metallic compounds that appear in OQMD but not in AFLOW (Methods), and scored them against OQMD, a second independent PBE database. After a per-element re-alignment fit on the scored compounds themselves (Methods J) and the same trimming as above, the same models lead on the 12,409 compounds whose elements carry unary references. eqV2 (15), eSEN (17), UMA (20), and SevenNet-MF (21 meV/atom) lead, as they do on the re-referenced AFLOW leaderboard, and the first-generation MP-trained models trail far behind (M3GNet 61, CHGNet 63). About two fifths of the scored compounds (4,821 of 12,409) have a structure that appears in none of the models' training sets. On those compounds every model is less accurate than on the 7,588 seen ones, by 1.3 to 17.7 meV/atom (1.3 to 14.9 for the models trained on OMat24 or Alexandria), and the ranking barely changes (Spearman $\rho$ 0.978 with the full set, with the top five in the same order; eqV2 15.8, eSEN 20.7, UMA 22.0 meV/atom). The test does not rule out chemical familiarity, because the elements and neighboring compositions are in training.

A direct test bounds memorization itself. On the OQMD compounds that only Alexandria contains, we compare how much the four models trained on Alexandria (MACE-MPA, eqV2, eSEN, SevenNet-MF) improve over never-seen compounds with how much the OMat24-only models, which never saw these compounds, improve. None of the three ways of matching composition detects a memorization advantage ($-$1.4 [$-$2.2, $-$0.6], $-$0.1 [$-$0.9, +0.6] and $-$1.2 [$-$2.3, $-$0.2] meV/atom), so the data bound any such advantage at under 1 meV/atom (the largest upper limit is 0.6). The same test does detect a training-set advantage where one exists. On Materials Project compounds, models trained only on the Materials Project improve by 5 to 15 meV/atom more than OMat24-only models (range over the three ways of matching composition; Supplementary Note 15). Having seen a compound in Alexandria therefore does not make a model predict it measurably better.

The alignment itself exposes a convention. Every one of the fourteen models needs the same $\sim$0.7 eV/atom shift on Tb compounds, an OQMD-vs-AFLOW reference convention that the alignment removes identically for all.

The complementary test scores the models on data they did train on. We ran all fourteen on 50,002 configurations from OMat24's held-out validation split (Methods) and scored forces only. OMat24's VASP settings differ from the Materials Project's by 13.5 meV/atom, more than the entire published energy-MAE range, so its energies cannot be compared without re-referencing. The models separate completely by training corpus. The eight OMat-trained models score 0.055--0.082 eV/\AA{} and the six MP-trained ones 0.135--0.254, and the closest pair across the divide is still a factor of 1.6 apart. This test measures accuracy on the training distribution itself, which the familiar/novel comparison cannot reach. The split holds out individual configurations, not whole materials. Distortions of the same crystal mostly sit in the training set, so each model has seen close relatives of everything it is tested on.

On this test UMA leads at 0.055 and eqV2 follows at 0.061, so the model that tops our independent-reference leaderboard does not top OMat24's own held-out data. As the data move toward its training distribution, eqV2's advantage does not grow, as a memorization advantage would. The order does depend on how many strongly distorted frames the set holds (Supplementary Note 15). Forces make this argument in a way energies could not, because no elemental reference convention affects them.

Supplementary Note 15 breaks this test down by reference force and by how the frames were generated.

The error is not spread evenly across the benchmark. Because AFLOW's libraries are combinatorial enumerations of prototypes, the formation-energy distribution of the structures peaks near zero but skews strongly toward unfavorable energies, with a median of +172 meV/atom and 71\% of structures above zero (Fig. 8).

We bin every structure by its distance above the hull and track each model's median error (Fig. 9). The relationship is tight, shared, and close to monotonic. Error stays low and flat for the near-hull structures, where the best models reach or fall below the 6--8 meV/atom floor, then rises by a factor of 1.6 to 3.3 toward the most metastable structures. The model ordering is preserved across the range. The same rise appears against raw formation energy, only less cleanly, since $E_{\mathrm{hull}}$ is the sharper axis. The rank correlation runs 0.84 to 0.99 for eleven of the fourteen models, and 0.87 to 0.99 for all fourteen once the sparse bins above +2 eV/atom are excluded (Supplementary Note 15; Supplementary Fig. 2).

\begin{figure*}[!tb]\centering
\begin{minipage}[c]{0.63\textwidth}\centering
\includegraphics[width=\linewidth,height=0.5\textheight,keepaspectratio]{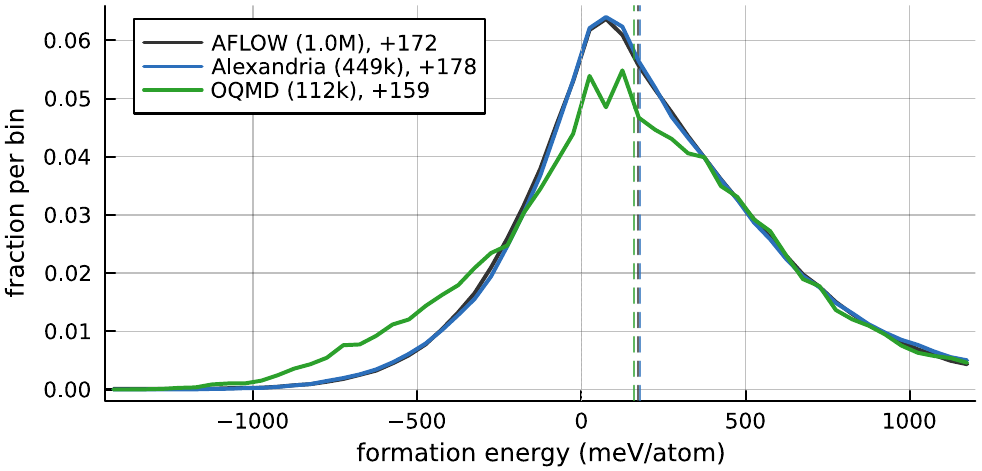}
\end{minipage}\hfill%
\begin{minipage}[c]{0.34\textwidth}
\caption{Formation-energy distribution of AFLOW's metallic benchmark (black, median +172 meV/atom, 71\% above zero), compared with the AFLOW structures that Alexandria also holds as the same crystal (blue, 449k; Alexandria's own energies for them put 72.1\% above zero, as do AFLOW's) and OQMD's \textbf{complete} metallic set (green, 111,901 entries, median +159). AFLOW's combinatorial libraries deliberately sample unfavorable structures, and all three databases show the same skew. (OQMD is shown in full, not only where it overlaps AFLOW, so that its curve shows how an independently built database samples metals.)}
\end{minipage}
\end{figure*}
Every model's error rises with distance above the hull, and two explanations compete. Far from equilibrium the PES is more anharmonic and near-degenerate, so these structures may be intrinsically harder. Even a familiar prototype is also rarely seen at such an unfavorable composition, so the training data may be less relevant there. This curve cannot separate the two, and the matched comparison below separates them only in part. The rise also depends on the sign of the energy, not only its size. At the same \textbar{}$E_f$\textbar{}, about 1 eV/atom, the models are 1.6 to 4.7 times worse (median 2.8$\times$) for unfavorable compounds ($E_f$ > 0) than for stable ones ($E_f$ < 0).

We separate familiarity from hull distance by comparing familiar and novel structures at matched $E_{\mathrm{hull}}$ and number of elements (Methods J). For the eight OMat-trained models, the gap falls only to 4.3--11.0 meV/atom (eqV2 from +5.3 to +4.3, eSEN from +11.6 to +10.3, UMA unchanged at +7.9). On the novel structures with a new composition in a familiar chemical system, the same matching takes the 12.3--22.1 meV/atom gap down to 2.9--10.4. That tier's gap came largely from where those structures sit, far above the hull, where every model is worse.

Over all novel structures, hull distance accounts for little of the OMat-trained models' gap (at most 20\%, and Orb-v3's gap grows slightly under matching). A residual of 4.3 to 11.0 meV/atom remains for every one of them, each with an interval that excludes zero. Once hull distance and the number of elements are matched, these models are thus 4 to 11 meV/atom less accurate on novel structures than on familiar ones. Matching on formation energy and the number of elements instead leaves 4.4--11.7 meV/atom. Balancing element content as well as hull distance and the number of elements leaves 6.8--10.4 meV/atom for seven of the eight and +2.7 [+2.1, +3.2] for eqV2 (Supplementary Table 20), so no control we tried removes the gap of any model trained on OMat24 or Alexandria. Supplementary Note 15 gives the gap under every matching we tried.

The five models trained without OMat24 or Alexandria need one more step. Their familiar sets are small (about 33,200 structures here), and 60\% of those structures are elements. Far above the hull the familiar side is therefore mostly high-energy elemental polymorphs, and matching on hull distance alone compares elements against compounds. For CHGNet that comparison even reverses the sign of the gap. When we match on hull distance and the number of elements together, CHGNet keeps 43 meV/atom, SevenNet-0 29, MatterSim-1M and M3GNet 12 each, and MatterSim-5M 10, every interval excluding zero. Matching removes a third to over two thirds of these gaps (35--71\%). A model trained without OMat24 or Alexandria loses 33 to 78 meV/atom on novel structures before matching, far more than an OMat-trained one. These models' remaining gaps are less certain. M3GNet's interval includes zero once element content is also balanced, as do SevenNet-0's and CHGNet's on the same-system subset (Supplementary Table 20).

As causes of the rise, we can rule out reference noise, and training sparsity only in part. Alexandria, a major training source, samples AFLOW's unfavorable metallic structures about as heavily as AFLOW does (Fig. 8). But coverage by Alexandria and the Materials Project falls with distance above the hull, from 84\% of structures within 25 meV/atom of it to 43\% at 400--1,600 meV/atom (Supplementary Note 15). Measured the same way, the DFT-vs-DFT floor stays flat in its 5--10 meV/atom band across the entire $E_{\mathrm{hull}}$ range (Fig. 1), so the rising error belongs to the models, not to the benchmark, and every model shows it.

\begin{figure*}[!tb]\centering
\includegraphics[width=\textwidth,height=0.5\textheight,keepaspectratio]{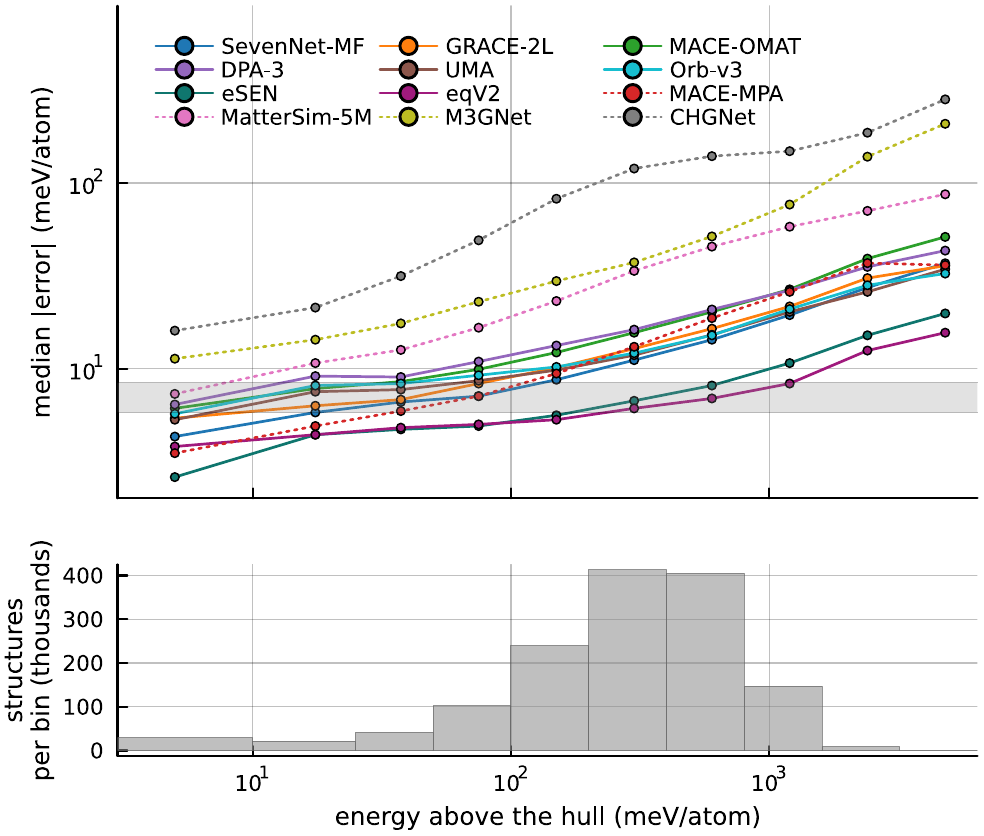}
\caption{Median model error versus AFLOW $E_{\mathrm{hull}}$ (top, log-log, twelve models), the direct ``distance above the hull'' form of the formation-energy binning (Supplementary Note 15). Color identifies the model, and line style the training generation. The shaded band is the 6--8 meV/atom median floor (noise-floor section). Error is lowest near the hull, where the best models reach or fall below that floor, and rises by a factor of 1.6 to 3.3 toward the most metastable structures. The lower panel gives the count per $E_{\mathrm{hull}}$ bin. Most structures sit well above the hull, and the near-hull bins are comparatively sparse.}
\end{figure*}
\begin{figure*}[!tb]\centering
\begin{minipage}[c]{0.63\textwidth}\centering
\includegraphics[width=\linewidth,height=0.5\textheight,keepaspectratio]{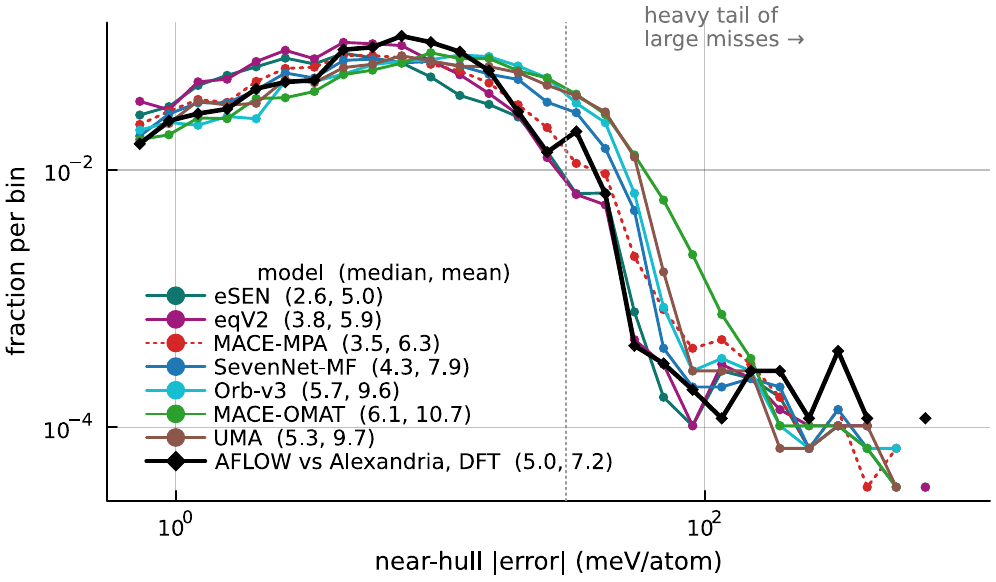}
\end{minipage}\hfill%
\begin{minipage}[c]{0.34\textwidth}
\caption{Near-hull disagreement (log-log): per-structure \textbar{}error\textbar{} for the 29,175 structures of the common set within 10 meV/atom of the AFLOW hull, the same for every model, with the DFT-vs-DFT yardstick in black ($\lvert\Delta E_f\rvert$ between AFLOW and Alexandria on the 25,546 of the same structures that Alexandria holds as the same crystal). The legend gives each curve's (median, mean) in meV/atom.}
\end{minipage}
\end{figure*}
Near the hull, where the typical error is smallest, the leading models are as close to AFLOW as a second DFT database is (Fig. 10). eSEN (median 2.6 meV/atom), MACE-MPA (3.5), and eqV2 (3.8) are lowest, and UMA, Orb-v3, and MACE-OMAT reach 5.3--6.1. The tails are alike, with each of these models missing by more than 100 meV/atom on 0.1--0.2\% of its near-hull structures. As an absolute yardstick, Figure 10 also shows the same distribution for AFLOW against Alexandria (median 5.0, mean 7.2 meV/atom). Every model is scored on the same 29,175 near-hull structures of the common set (Methods G), and the yardstick on the 25,546 of them that Alexandria also holds, so the models and the yardstick are measured on nearly the same population. Even the two DFT databases disagree by more than 100 meV/atom on a few of these structures (36 of 25,546), the same bad-reference class the bad-references section examines.

The degradation on novel structures is real for every model, and for the OMat-trained models it is moderate, 5 to 12 meV/atom before matching and 4 to 11 after. A cohort that had merely memorized its training set would not stay within 5 to 12 meV/atom of its familiar-structure error on a set that is about half or more novel for every model. The loss of accuracy is also smallest where accuracy matters most. Within 100 meV/atom of the hull, among the thermodynamically plausible phases a discovery campaign actually pursues, novel structures are predicted only 0.5 to 3 meV/atom worse than familiar ones. The loss grows to 6 to 20 meV/atom only 400 meV/atom or more above the hull, among strongly metastable structures a screen would discard anyway.

\section{Discussion}
\begin{figure*}[!tb]\centering
\includegraphics[width=\textwidth,height=0.5\textheight,keepaspectratio]{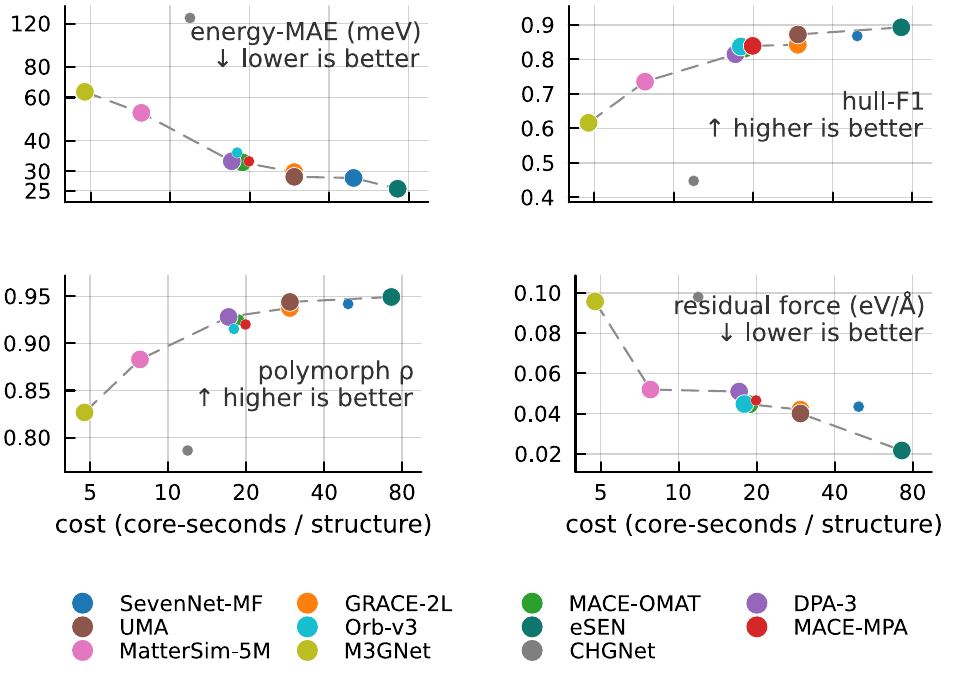}
\caption{Accuracy versus compute cost (median core-seconds per structure, log x) for eleven headline models, one panel per metric. In each panel a large dot marks a model on that metric's Pareto frontier (dashed). Smaller dots are dominated. The efficient set shifts with the metric. Orb-v3 and UMA hold the mid-cost frontier for stability and polymorph ordering, joined by MACE-OMAT and GRACE-2L for forces, while DPA-3, MACE-OMAT, GRACE-2L, UMA, and SevenNet-MF carry it on raw energy. DPA-3 sits on all four frontiers, eSEN holds the high-accuracy tip of all four, and CHGNet is dominated throughout. The energy panel uses a log scale, since CHGNet's 127 meV/atom would otherwise compress the 25--36 band where the other models separate. eSEN's cost is a matched CPU re-measurement, and eqV2 is omitted because its direct-force head barely relaxes (Methods; leaderboard section).}
\end{figure*}
The apparent failures of the leaderboard through generalization sections sort into five kinds a benchmark must tell apart: a reference convention, an error in the reference itself, an irreducible disagreement between two DFT codes, an artifact of the benchmark's own analysis, and a genuine model limitation. The third kind is what the floor measures, where beating it would mean fitting one DFT code's settings, not a better PES. The magnetic 3d discrepancies are of the first two kinds, a per-element offset or an error in one database's entry (for Alexandria, usually a missed magnetic state). The +U shear cases are a mismatch between the Hamiltonian of the reference and that of the training data, which no floor pair measures, since none carries +U.

Only the fifth kind, a genuine model limitation that neither re-referencing nor a protocol control can remove, can be charged to the model. A disagreement earns that label only after a reference error and a protocol mismatch have both been ruled out. Table 5 gathers the controls that do the ruling out.

Below the leaders, the rankings on forces, elastic moduli, and symmetry retention differ from the ranking on energy, while energy and stability largely agree (reference-free section). No single leaderboard therefore stands in for the rest. Multi-metric reporting is not new (Matbench Discovery already reports several statistics of its stability task). What is new here is that these axes measure different properties and are scored against a reference no model trained on (Supplementary Note 8 gives every model on every metric).

Cost is an axis of its own, and it interacts with the choice of metric (Fig. 11). The efficient frontier shifts from panel to panel, so the cost-effective model depends as much on the budget and the target property as on any single leaderboard.

\begin{figure*}[!tb]\centering
\begin{minipage}[c]{0.63\textwidth}\centering
\includegraphics[width=\linewidth,height=0.5\textheight,keepaspectratio]{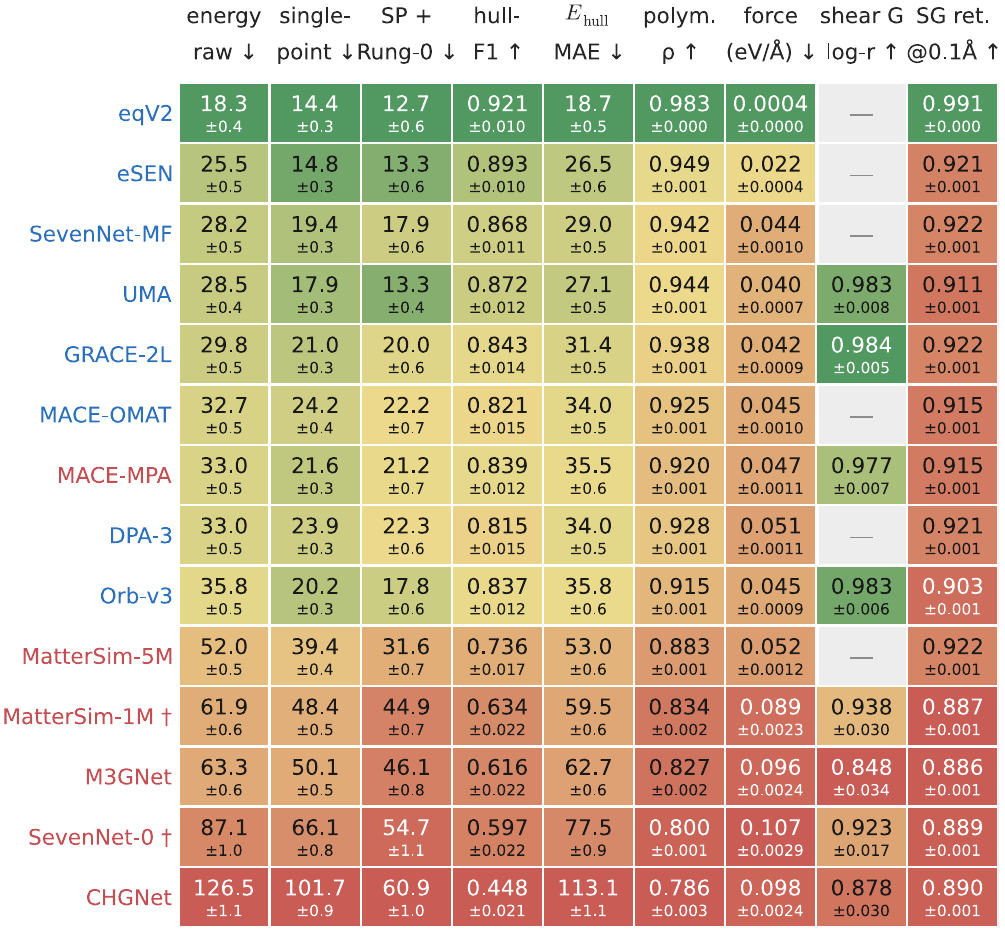}
\end{minipage}\hfill%
\begin{minipage}[c]{0.34\textwidth}
\caption{The full leaderboard: fourteen models (rows, sorted by raw formation-energy MAE; blue OMat / red MP; \textdagger{} superseded within-family checkpoints) on nine axes, each column color-graded independently from best (green) to worst (red). Every cell carries its 95\% bootstrap half-interval, resampled over chemical systems. Gray dashes mark metrics not run for a model (shear: the eight elastic-task models, subset without +U; Methods). The generalization gap is not a column here; Fig. 7 gives it. From column to column, the order changes most below the leading models, as in Fig. 5.}
\end{minipage}
\end{figure*}
On most of those axes (Fig. 12), the strongest predictor of where a model lands is its training set. The OMat-trained models lead the energy ranking and the reference-free metrics, degrade least on unseen structures, and share their errors. Two models trained on the same corpus make more similar errors than two trained on different corpora (mean Spearman $\rho$ 0.61 versus 0.40 over the 45 pairs of ten models; no other way of assigning four of the ten to the MP group gives as large a gap, p = 1/210 = 0.005; Methods I). Readers can add their own model to Fig. 12. The scripts in \texttt{scripts/\allowbreak byom/\allowbreak } run any model with an ASE calculator through the same protocols and redraw the figure with it included, on a subset that runs on a workstation or on the full benchmark (Code availability).

The panel's one same-recipe pair tests this claim. eSEN and eqV2 were trained on the same corpora (OMat24 pretraining, MPtrj + sAlex fine-tune) but differ in architecture, scale, and force objective. With both controls applied, their energy gap closes to 0.6 meV/atom (leaderboard section), so the raw gap came mostly from relaxation drift. eqV2 still leads eSEN on hull-F1, by 0.028 after relaxing (0.921 versus 0.893) and narrowly at AFLOW's geometry (0.930 versus 0.923). The data do not say whether architecture or closeness to AFLOW's protocol is the reason. On polymorph ordering the two tie at AFLOW's geometry, so eqV2's relaxed lead there is a relaxation effect. The OMat24 test agrees. Within the OMat tier, architecture spreads the force error only 1.5-fold, compared with a gap of at least 1.6-fold between tiers, and UMA, not eqV2, leads there. For model developers, training data has the larger effect, but it is not the only one.

Judging the next generation well takes three inexpensive habits. Score against at least one DFT reference the models were not trained on, so the test measures transfer rather than recall. Report the error before and after re-referencing, since the difference bounds how much of it a convention mismatch could explain. Finally, compare with a measured DFT-vs-DFT floor, quoted with the same trimmed mean as the model error and on the same structures, so a 7 meV/atom gap is not mistaken for skill when two careful codes disagree by that much on typical structures.

Re-referencing also matters to anyone who uses these potentials. A per-element shift cancels exactly in any comparison at fixed composition, such as polymorph ordering, so for those uses the re-referenced error is the one a user meets. It does not cancel when a model's energies are compared with a database's. A candidate relaxed with a potential and judged against the Materials Project's or AFLOW's convex hull mixes two conventions. The databases themselves re-reference routinely \cite{stevanovic2012,kirklin2015,wang2021b}, and a user can remove the same mismatch the same way, with one constant per element fit to a set of DFT calculations in the chemical system of interest. For the three models trained without Alexandria or OMat24 (MatterSim-5M, M3GNet, CHGNet) that step removes 7 to 54 meV/atom; for the other nine, 1 to 3. Rung-0 also tells a user which remedy a model needs. An error that is mostly per-element is removed by a few constants. An error that survives re-referencing is structure-dependent, either a PES error or a protocol difference of the kind the floor measures, and only fine-tuning or better training data reduce it.

The same caution applies to ``novelty.'' On databases this large, a structure a model has never seen is still almost always close in composition and chemistry to ones it has, so a truly out-of-distribution test is nearly impossible to construct. The generalization we can actually measure is transfer to an independent DFT reference, not escape from the training distribution.

Our benchmark and Matbench Discovery agree at tier level (the models atop Matbench Discovery are the models atop this leaderboard) and disagree exactly where our floor puts the limit of a transferable ranking. Matbench Discovery's leader, eSEN \cite{fu2025}, and its same-recipe sibling eqV2 swap places here.

The benchmark has limits. It uses one reference database (AFLOW) and one functional (PBE), so every conclusion is about agreement with raw-PBE DFT, and we drop the elemental entries AFLOW computed otherwise (benchmark section and Methods). It covers only ordered metallic structures (benchmark section), except for the elastic set, which includes nonmetals (reference-free section). The shear test cannot separate +U from element identity or correlated-electron effects (reference-free section). Of the discrepancies in the bad-references section, 7,067 have no same-crystal Alexandria entry, so their cause is unsettled. Rung-0 bounds the effect of a convention rather than measuring it, because it cannot tell a convention from an element-wise bias (leaderboard section). Familiarity is tagged against public stand-ins for each model's training data, some of which hold more than a model saw, and it depends on the tolerances of the structure match (generalization section; Methods J). The elastic task ran on the original eight-model panel (Methods A).

The next generation therefore has a clear, narrow target. Re-referenced, the best model sits 3 meV/atom below the floor, eSEN 2 above it, and UMA, SevenNet-MF, and GRACE-2L 5 to 7 above it. What still separates the models lies far above the hull. Every model's error rises with distance above the hull (Fig. 9), where training coverage thins out, so that is where targeted data on an independent reference could help most. As the models improve past the floor, a gain will be visible only if the comparison follows the three habits above and controls the protocol before crediting the model. We hope the controls assembled here make that easier for the benchmarks that follow.

\textbf{Data availability.} Everything the analysis reads is archived at Zenodo (DOI: 10.5281/zenodo.21301563): the per-structure benchmark records for every structure--potential pair, stored as JSONL with one row per pair keyed by AFLOW AUID and model, carrying energies, relaxation metrics, and space groups; the matching records for the elastic subset, the symmetry-tolerance sweep (which also stores every model's relaxed structure), and the OQMD cross-database test; the per-structure formation-energy table derived from them; the test-set manifest; the same-crystal match tables (every StructureMatcher match from the exhaustive search, with the space-group test applied) behind the noise-floor and overlap analyses; and the prototype-fingerprint tables they are compared with. AFLOW, Alexandria, OQMD, and Materials Project data were obtained from their public APIs and bulk downloads (AFLOW via the AFLUX REST API), and OMat24 from its release \cite{barrosoluque2024}. The exact queries and the versions of the bulk downloads are recorded in the archive.

\textbf{Code availability.} All analysis and figure-generation code, per-model environment lockfiles, and the model-checkpoint provenance are archived in the same Zenodo record (DOI: 10.5281/zenodo.21301563). Every data figure regenerates from the summary tables in the archive with the provided scripts, and every value in Tables 2--4 is read or computed from them. The scripts that compute those summary tables from the per-structure data are included. Several of them need cluster-scale resources. The scripts in \texttt{scripts/\allowbreak byom/\allowbreak } benchmark any model with an ASE calculator under the same protocols and redraw Fig. 12 with it included, on a stratified subset that runs on a workstation or on the full benchmark.

\textbf{Funding.} The authors received no funding for this work.

\textbf{Author contributions.} G.L.W.H. conceived and designed the benchmark and its controls-first methodology (the independent-DFT reference, the DFT-vs-DFT noise-floor calibration, and the held-out per-element re-referencing, Rung-0), and led the relaxation-fidelity, elastic, and +U analyses, the analysis and figure pipeline, and the original draft. J.N.P. led the convex-hull stability and training-overlap generalization analyses, contributed the leaderboard-table figure, drafted portions of the Methods and Supplementary Information, and provided an extensive section-by-section review of the manuscript. Both authors interpreted the results, reviewed and edited the manuscript, and contributed equally to this work.

\textbf{Competing interests.} The authors declare no competing interests.

\bibliographystyle{unsrtnat}
\bibliography{refs}

\clearpage
\section*{Methods}
The subsections trace the pipeline in the order it runs: build the test set, relax, reference, score, then apply each control. Every ``(Methods)'' pointer in the main text lands on the procedure that produced its number.

\subsection*{A. The benchmark: reference, test set, and potentials}
\textit{The reference} is three parts of AFLOW \cite{curtarolo2012,calderon2015}: the binary (LIB2) and ternary (LIB3) prototype libraries for the ordered intermetallics that are our core test, the unary entries for elemental reference states, and the Automatic Elasticity Library (AEL) for the DFT elastic tensors of Methods H \cite{toher2017}. For each structure we take AFLOW's DFT-relaxed geometry and its formation enthalpy per atom (\texttt{enthalpy\_\allowbreak formation\_\allowbreak atom}) as ground truth. AFLOW's LIB2/LIB3 alloy libraries carry no post-hoc correction (raw PBE). We access AFLOW through its AFLUX REST API \cite{rose2017}, keyed by each entry's AFLOW unique identifier (AUID). On the exhaustive fetch, the inline geometry matched the CONTCAR to within 0.002\%.

\textit{The test set} has a metals-only core, and to enter it a structure must pass four filters: every element in the 46-element \texttt{METALS} set of \texttt{scripts/\allowbreak build\_\allowbreak testset.\allowbreak py} or one of twelve metalloids/nonmetals (B, C, N, O, Si, P, S, Ge, As, Sb, Se, Te); fewer than half its atoms nonmetals; an arity (number of elements) of one to three and a reduced cell of at most 40 atoms (60 where AFLOW elastic data exist); and no Hubbard U. Hubbard U enters only the labeled extended strata and the lanthanide probes: all of E2 (19,882) and its elastic subset (2,749), 4,857 of the 9,153 E3 structures, 105 of the 884 lanthanide unaries, and one lanthanide alloy, 1.9\% of the benchmark in all. No energy metric includes a +U structure. Within that scope we take LIB2 and LIB3 essentially in full: $\sim$1.45M ordered structures (1,454,458)---$\sim$1.38M binary/ternary decorations at the core (202,724 LIB2, 1,178,736 LIB3), plus the elemental unaries that supply references ($\sim$29k), a lanthanide probe (E3, 9,153), lanthanide unaries (ln\_unary, 884), lanthanide alloys (ln\_alloy, 3,189), the labeled extended-chemistry strata, the AEL elastic subset, and a 32-structure quaternary set (AFLOW's alloy libraries stop at ternary). The extended strata are never pooled into headline metrics. E1 holds standard-DFT (no-U) nonmetal compounds, E2 holds +U entries admitted for geometry only (AFLOW reports no formation enthalpy for any +U intermetallic), and E3 is the lanthanide probe (lanthanide intermetallics of any U status). The lanthanide unary and alloy strata (ln\_unary, ln\_alloy) admit only the trivalent lanthanides (La, Pr, Nd, Gd, Tb, Dy, Ho, Er, Lu; mixed-valent Ce, Sm, Eu, Tm, Yb excluded). Actinides are excluded. We also drop the 1,575 elemental entries that AFLOW computed without plain PBE (LDA, SCAN, older GGA pseudopotentials, RPBE, and one spin-orbit Bi), whose energies are not on the PBE scale (\texttt{results/\allowbreak analysis/\allowbreak unary\_\allowbreak nonpbe\_\allowbreak exclusions.\allowbreak csv}; Supplementary Note 2). Sampling is monotone: each stratum's entries are taken in ascending md5 of the AUID up to a cap raised over five waves, so each wave is a strict superset of the last (wave caps in \texttt{scripts/\allowbreak build\_\allowbreak testset.\allowbreak py}, superset selection in \texttt{src/\allowbreak umlip\_\allowbreak bench/\allowbreak sampling.\allowbreak py}). The same ordering partitions the work across cluster jobs, and the runner skips any structure--potential pair already on disk, so the campaign restarts cleanly across shards, waves, and resubmissions.

\textit{The potentials} of Table 1 joined the panel in three waves: an original eight-model panel ran the full production campaign including the elastic task, four models joined mid-campaign, and eSEN and eqV2 arrived last. All fourteen run the OQMD cross-check of Methods J. Each model ships an ASE calculator in its own pinned virtual environment, which isolates incompatible dependency stacks (TensorFlow for GRACE-2L, PyTorch/fairchem for the rest). Checkpoint identifiers, package versions, and loaders are in \texttt{docs/\allowbreak model\_\allowbreak provenance.\allowbreak md} (Supplementary Note 1).

\subsection*{B. Relaxation and formation energies}
\textit{Relaxation} always starts from AFLOW's DFT-relaxed geometry (rationale in the benchmark section), with an identical protocol across all fourteen potentials: ASE FIRE with a FrechetCellFilter relaxing cell and positions together, maximum force 0.02 eV/\AA{}, capped at 500 steps \cite{bitzek2006,larsen2017}. No symmetry is imposed, so space-group retention (Methods G) is a measured outcome. A guard aborts and records a per-structure failure if the cell volume leaves the window 0.3--6.0$\times$ its start or any interatomic distance drops below 0.5 \AA{} (this also converts GRACE-2L's occasional collapsing-cell neighbor-list blow-up into a recorded failure rather than a node crash). Geometry fidelity is scored against AFLOW with pymatgen's StructureMatcher \cite{ong2013} (ltol 0.3, stol 0.5, angle\_tol 10$^\circ$, no supercell attempts) and with space groups from pymatgen's SpacegroupAnalyzer (which wraps spglib \cite{togo2018}) at symprec 0.01 \AA{}. A post-relaxation interatomic distance below 0.7 \AA{} marks a structure collapsed, distinct from the 0.5 \AA{} abort guard.

\textit{Formation energies} use the model's own elemental references: $E_f = E/\mathrm{atom} - \sum_{\mathrm{el}} x_{\mathrm{el}}\,E_{\mathrm{ref}}(\mathrm{el})$, with $x_{\mathrm{el}}$ the atomic fraction and $E_{\mathrm{ref}}(\mathrm{el})$ the minimum relaxed energy per atom over the element's unary stratum, without the non-PBE entries of Methods A. Candidate ground states (fcc, bcc, hcp, and ASE's default bulk reference) are seeded at the element's reference atomic volume (the per-atom volume of ASE's bulk crystal, or a covalent-radius fcc volume where unavailable, e.g. Mn, Ga) and then fully cell-relaxed, so the seed volume is only a starting point (\texttt{src/\allowbreak umlip\_\allowbreak bench/\allowbreak references.\allowbreak py}; production references in \texttt{analyze.\allowbreak py}). Each molecular reference (H, N, O, F, Cl) is a diatomic molecule in a box, relevant only at the extended-chemistry edges. An element whose per-potential reference fails to converge is held on a per-model list, and its structures are dropped for that model on both sides of every comparison, which is why some cross-database subsets (Methods J) shrink to compounds whose elements all carry a converged reference.

\textit{Elemental-reference screening} is needed because the reference is a minimum (benchmark section). A structure can be a reference only if it passes three clauses, applied per model and element and identically to all fourteen models (\texttt{scripts/\allowbreak screen\_\allowbreak unary\_\allowbreak refs.\allowbreak py}, projected onto \texttt{results/\allowbreak analysis/\allowbreak unary\_\allowbreak blocklist.\allowbreak csv} by \texttt{scripts/\allowbreak build\_\allowbreak unary\_\allowbreak blocklist.\allowbreak py}). S0 rejects a one-atom cell without an AFLOW space group (AFLOW's isolated-atom references, one atom in a 15 \AA{} cube). S1 rejects any low-energy group that sits more than 500 meV/atom below the rest of that element's structures and holds at most 20\% of them. S2 rejects a structure whose relaxed volume differs from AFLOW's by more than 25\% with its space group retained. The 500 meV/atom gap lies between the largest physical polymorph gap (250 meV/atom; W, Mo, Cr, Nb) and the smallest pathological one (942 meV/atom, MACE-MPA Fe). The 25\% bound lies between the $\alpha$-Sn control ($-$1.8\% under MACE-MPA, $-$0.3\% under eSEN) and the collapsed diamond-Fe basin ($-$29\% to $-$31\%), and raising it anywhere below 29\% moves no reference by more than 3 meV/atom. The screen rejects 4,500 structures (3,015 by S0 and 1,485 by S2; S1 rejects none beyond these). It moves ten references by more than 10 meV/atom, across five models (\texttt{results/\allowbreak analysis/\allowbreak unary\_\allowbreak ref\_\allowbreak changes.\allowbreak csv}): DPA-3's Gd (+1.02 eV/atom, structures whose volume grows by 59--87\%), MACE-MPA's Fe (+0.94, a collapsed diamond cell), and, set by isolated atoms, UMA's Pd (+0.34) and eSEN's Ti, Sn, V, and Ga (+0.15 to +0.29). The rest (UMA's and Orb-v3's Cs, eSEN's Be) move by 10--84 meV/atom. The screen replaces a hand-curated list of 189 structures, which had already removed eSEN's Cs structure (+44\% volume, 3.6 eV/atom below its sound Cs structures).

\subsection*{C. The single-point control}
The single-point control scores every model at AFLOW's geometry, from the energy each relaxation recorded before its optimizer's first step (\texttt{e\_\allowbreak initial\_\allowbreak per\_\allowbreak atom}), with no re-relaxation and no new model evaluations (\texttt{scripts/\allowbreak singlepoint\_\allowbreak control.\allowbreak py}). It costs nothing, since the optimizer must evaluate the starting point anyway. It uses the same elemental references and the same 1 eV/atom trim as the relaxed column, with the trim applied to the \textit{relaxed} error so both columns share one row set. The difference between the columns is then the geometry effect alone. As a check, eqV2's median per-structure \textbar{}single-point $-$ relaxed\textbar{} difference is 0.0 meV/atom: its direct-force head reads AFLOW's minima as already converged, and 65\% of its structures take no optimizer step, compared with 3--5\% for eSEN and UMA. Its relaxation drift is 3.9 meV/atom, compared with about 10.6 for those two. That 3.9 meV/atom gap between its two Table 3 columns comes from the 35\% in which its relaxation did move.

\subsection*{D. The trimmed mean and its uncertainty}
\textit{The trimmed mean} is our error statistic. The per-structure error is $E_f(\mathrm{model}) - E_f(\mathrm{AFLOW})$ per atom, and we report a \textit{trimmed} mean absolute error, the mean of \textbar{}error\textbar{} after setting aside the few structures whose \textbar{}error\textbar{} exceeds 1 eV/atom, which we count and report separately rather than discard. We did not tune the 1 eV/atom cutoff. A sensitivity sweep checks the cutoff (Supplementary Note 6). We apply the same trim to the model errors and to the DFT-vs-DFT floor (Methods E), so both sides of every comparison use the same trimmed mean.

\textit{Uncertainty} comes from resampling whole chemical systems, not individual structures, so near-duplicate decorations do not make the intervals look narrower than they are. Each metric is computed once per system, then bootstrapped by resampling systems with replacement (B = 2000, percentile 2.5/97.5, fixed seed; \texttt{scripts/\allowbreak bootstrap\_\allowbreak ci\_\allowbreak all.\allowbreak py}).  Hull, polymorph, and energy metrics all resample the same unit (the chemical system). Hull and polymorph metrics use the common set C of Methods G. Energy metrics use every row that passed the trim.

\textit{Conventions used throughout} the paper are these. In Table 3 and wherever the text quotes an energy MAE, every energy metric uses all benchmark structures with an AFLOW formation enthalpy and a model energy, about 1.42M per model,  after the 1 eV/atom trim. A model error is quoted before re-referencing unless labeled after Rung-0 or re-referenced.  ``All fourteen'' means every model of Table 1, and the twelve headline models omit the two superseded checkpoints (SevenNet-0 and MatterSim-1M, marked \textdagger{}). The consensus test of the bad-references section uses the original eight-model panel (CHGNet, MACE-MPA, M3GNet, SevenNet-0, Orb-v3, MatterSim-1M, GRACE-2L, UMA), and the six MP-generation models are the six MP rows of Table 1. The five models of the generalization section trained without OMat24 or Alexandria are those six less MACE-MPA, which also trained on an Alexandria subset, and none of the five trained on OMat24 or Alexandria. The four models trained on the Materials Project and Alexandria are MACE-MPA, eqV2, eSEN and SevenNet-MF. DPA-3 trained on OMat24 and MPtrj but, of Alexandria, only on its two-dimensional subset, so it is tagged against OMat24 and the Materials Project alone (Methods J). Its other training sets were not matched against AFLOW, so its familiar fraction is a lower bound. The four OMat24-only models are GRACE-2L, MACE-OMAT, Orb-v3 and UMA. AFLOW entries were fetched through AFLUX in five waves between 11 and 17 June 2026 (cluster logs \texttt{logs/\allowbreak testset\_\allowbreak build*.\allowbreak log}, \texttt{wave4\_\allowbreak fetch.\allowbreak log}). Alexandria energies come from release 2024.12.15, OQMD's from the qmpy v1.8 database dump, and the Materials Project's from a June 2026 query. Structure matching against Alexandria uses the 2024.12.15 structures, the release the energies come from. Only the OMat24 parent structures absent from that release come from release 2025.07.02 (Methods J). The medians of the floor and of the leader check use all pairs, before the trim, while the binned model medians of Figs. 9 and 10 use the rows of the hull metrics.

\subsection*{E. The DFT-vs-DFT noise floor}
\textit{Matching} treats two structures from different databases as the same crystal if pymatgen's \texttt{StructureMatcher} \cite{ong2013} matches them directly at its default tolerances (ltol 0.2, stol 0.3, angle\_tol 5$^\circ$, primitive cells, no supercell), in the sense that \texttt{get\_\allowbreak rms\_\allowbreak dist} returns a mapping, and their space groups, assigned by pymatgen's SpacegroupAnalyzer at symprec 0.1 \AA{}, are the same. The search is exhaustive within each composition. We compare every pair of structures with the same reduced formula across AFLOW, Alexandria, the Materials Project, OQMD, and the Alexandria parents of the OMat24 training rows, 16.77 million pairs, and the matcher completes every one (\texttt{scripts/\allowbreak smfam\_\allowbreak build\_\allowbreak stores.\allowbreak py}, \texttt{scripts/\allowbreak smfam\_\allowbreak match.\allowbreak py}, \texttt{scripts/\allowbreak smfam\_\allowbreak smsg.\allowbreak py}). AFLOW's own XtalFinder \cite{hicks2021} compares structures against AFLOW's prototype library \cite{mehl2017,hicks2019,hicks2021b}; we use pymatgen throughout so that one tool compares structures from every database. The same-crystal matches between two databases join their structures into connected components. For the floor, each component gives one pair, its lowest-$E_f$ AFLOW structure and that structure's lowest-$E_f$ direct partner in the other database (\texttt{scripts/\allowbreak floor\_\allowbreak pairs\_\allowbreak smsg.\allowbreak py}). The space-group requirement removes 100,067 of the 815,539 AFLOW--Alexandria matches. A prototype fingerprint (the reduced formula, the space-group number, and the occupied Wyckoff positions) is a cheaper lookup but a different test. Of the 448,532 metallic AFLOW--Alexandria pairs it would have found 321,725 (72\%), missing mostly the same crystal described with different Wyckoff labels, and it proposes 17,344 pairs (5.1\% of its 339,202) that are not the same crystal. For AFLOW--OQMD, OQMD--Alexandria, and AFLOW--MP it would have found 42\%, 56\%, and 72\% of the pairs, and 0.7\%, 0.5\%, and 2.2\% of its own pairs are not the same crystal (\texttt{scripts/\allowbreak floor\_\allowbreak smsg\_\allowbreak vs\_\allowbreak fingerprint.\allowbreak py}). At tolerances (ltol/stol/angle\_tol) of 0.1/0.2/3$^\circ$, 0.2/0.3/5$^\circ$ (the default, used throughout), 0.3/0.5/10$^\circ$, and 0.4/0.6/15$^\circ$, the AFLOW--Alexandria trimmed MAE is 15.4, 16.4, 19.2, and 20.9 meV/atom, and the median stays between 5.7 and 6.1 (\texttt{results/\allowbreak analysis/\allowbreak floor\_\allowbreak smsg\_\allowbreak summary*.\allowbreak csv}).

\textit{Settings} differ less than one might expect. AFLOW and Alexandria report raw-PBE formation energies on metals. Alexandria applies the Materials Project's corrections, but they are zero on every metallic entry. OQMD fits the chemical potentials of 13 elements to experiment \cite{kirklin2015}. Four of them are metals in our set (Na, Ti, Sn and Hg), and 20\% of the AFLOW--OQMD floor pairs contain one of the four. On those pairs the fit shifts OQMD's formation energies by a median of $-$23 meV/atom. Removing the shift would lower the AFLOW--OQMD floor from 8.4 to 7.4 meV/atom (median) and from 25.8 to 22.6 (trimmed MAE), and the Alexandria--OQMD floor from 7.8 to 7.0 and from 29.9 to 27.0 (\texttt{scripts/\allowbreak f305\_\allowbreak oqmd\_\allowbreak offset.\allowbreak py}). We report the floor with OQMD's energies as published. Supplementary Note 4 gives each database's settings.  No floor pair carries +U (\texttt{scripts/\allowbreak ldau\_\allowbreak floor\_\allowbreak sm.\allowbreak py}, \texttt{results/\allowbreak analysis/\allowbreak ldau\_\allowbreak floor\_\allowbreak sm.\allowbreak csv}; every population's +U fraction is in Supplementary Table 6).

\textit{Median, trimmed mean, and checks} together define the floor. We report it two ways, each computed exactly as the model error is: the median $|\Delta E_f|$ and the trimmed MAE under the same 1 eV/atom trim (Table 2). Every pair is restricted to metallic prototypes, by the same nonmetal-exclusion set used throughout (Supplementary Note 5). Two checks establish that the floor is not an easy artifact. First, we fit a composition-linear per-element offset between each pair (result in the noise-floor section). We did not attempt richer structure-level corrections. Second, to rule out an easier structure set than the floor's, we recompute the leader's error on exactly the floor-defining structures, one per same-crystal component (\texttt{scripts/\allowbreak floor\_\allowbreak structure\_\allowbreak matched\_\allowbreak downstream.\allowbreak py}; results in the leaderboard section). Adding a volume criterion (\textbar{}$\Delta$V/V\textbar{} $\le$ 10\%) lowers the AFLOW--Alexandria floor further, to 5.7 and 13.3. We do not apply it, because a volume difference between two relaxations of the same crystal is part of the protocol disagreement the floor measures.

\textit{Materials Project} needs its own cut. An overlap cut that merely looks metallic is still nearly a third oxides and nitrides, and its removable gap is an oxide effect, so the MP row of Table 2 uses a clean metallic cut (results in the noise-floor section).

\subsection*{F. Rung-0: a held-out per-element re-referencing}
Rung-0 fits one energy constant per chemical element \cite{stevanovic2012} and subtracts it, composition-weighted, from the model's formation energies. The constants minimize the absolute error, the statistic we score (a least-absolute-deviation fit). The fit uses the element-fraction design matrix with a small ridge term ($\lambda$ = 10\textsuperscript{--}\textsuperscript{3}, needed because the fractions sum to one and the matrix is collinear), restricted to the metallic alloy strata (LIB2/LIB3, elemental and lanthanide unaries, lanthanide alloys, and the quaternary set) and to rows with \textbar{}error\textbar{} < 500 meV/atom. We evaluate the corrected errors with the same 1 eV/atom trim as the raw leaderboard. The fit is solved by iteratively reweighted least squares (weights 1/max(\textbar{}r\textbar{}, $\varepsilon$), with $\varepsilon$ lowered from 1 to 10\textsuperscript{--}\textsuperscript{4} meV/atom). We confirm that it is a minimum: neither moving any single constant by $\pm$0.01 or $\pm$0.1 meV/atom nor substituting the least-squares, Huber, or one-constant fits lowers the absolute error (\texttt{scripts/\allowbreak rung0\_\allowbreak lad.\allowbreak py}).  Supplementary Note 7 compares it with the least-squares fit of our first version and with a Huber fit.

The constants are fit on 70\% of the chemical systems and scored on the other 30\%, with no chemical system in both (\texttt{scripts/\allowbreak rung0\_\allowbreak alloy.\allowbreak py}, split materialized by \texttt{scripts/\allowbreak rung0\_\allowbreak hash\_\allowbreak split.\allowbreak jl}); Supplementary Note 7 gives a five-fold check. Rung-0 removes anything linear in composition, so its gain is a bound on what any change of elemental reference could remove (leaderboard section). We correlate CHGNet's offsets with the per-element MP$-$AFLOW and Alexandria$-$AFLOW reference differences over the 47 of its 49 fitted elements with at least 20 supporting structures in both fits (\texttt{results/\allowbreak analysis/\allowbreak rung0\_\allowbreak chgnet\_\allowbreak diagnosis.\allowbreak csv}; results in Supplementary Note 7).

Two derived columns follow. ``SP + Rung-0'' fits Rung-0 on the single-point errors, again held-out, so neither drift nor any per-element offset remains. It uses the alloy strata, trimmed on their own held-out single-point errors. ``Above floor'' is the model's after-Rung-0 trimmed MAE minus the AFLOW--Alexandria trimmed MAE, with both terms computed on the same structures. These are the held-out structures that define the floor, namely the AFLOW structures of the same-crystal AFLOW--Alexandria pairs of Methods E that fall in the 30\% of chemical systems held out of the Rung-0 fit, 125,455--125,947 per model. Rung-0 is fitted exactly as for the ``After Rung-0'' column. On these structures the floor is 16.7 meV/atom (\texttt{scripts/\allowbreak floor\_\allowbreak sm\_\allowbreak rung0.\allowbreak py}, \texttt{results/\allowbreak analysis/\allowbreak floor\_\allowbreak sm\_\allowbreak rung0.\allowbreak csv}).  The column's interval resamples the held-out chemical systems and recomputes the model's and the floor's trimmed MAE on the same resampled structures (B = 2000), with the per-element constants held at their fit. Repeating the comparison at the strictest and loosest matching tolerances of Methods E gives the tolerance columns of Supplementary Table 8 (\texttt{scripts/\allowbreak floor\_\allowbreak sm\_\allowbreak rung0\_\allowbreak ci.\allowbreak py}).

\subsection*{G. Reference-free metrics}
The hull and polymorph metrics are scored against AFLOW on one common set C for every model: the 1,407,872 structures that all fourteen models score, with both the relaxed and the single-point error within 1 eV/atom (\texttt{scripts/\allowbreak reffree\_\allowbreak singlepoint.\allowbreak py}, \texttt{scripts/\allowbreak reffree\_\allowbreak common\_\allowbreak regen.\allowbreak py}). We do not keep, for each model, only the structures it relaxed without a change of space group. That filter would remove exactly the structures a model relaxes away from AFLOW's and give each model a different set (1.41 million for eqV2, 1.18 million for UMA), so it would select on the outcome. The single-point columns score the same set from the energies at AFLOW's geometry (Methods C), where no structure can lose its symmetry. The Alexandria ceiling uses every AFLOW structure with an AFLOW energy, so no model enters it.

\textit{Convex-hull stability (hull-F1)} uses hulls we build ourselves. AFLOW provides its own hull analysis, AFLOW-CHULL \cite{oses2018}; we build every hull with one solver so that AFLOW's and each model's hulls are built identically. For each chemical system we build the formation-energy hull twice---from AFLOW's energies and from the model's---and label a structure stable if it lies within 25 meV/atom of the hull (hull-F1 as defined in the reference-free section). We also report the MAE of the energy above the hull. We compute the energy above the hull with a direct scipy lower-convex-hull solver \cite{virtanen2020,barber1996} over each structure's whole chemical space, so a ternary is measured against the binary phases of its sub-systems as well as the other ternaries (\texttt{scripts/\allowbreak hull\_\allowbreak proper.\allowbreak py}). Each element's corner is its AFLOW reference energy, $E_f = 0$. An elemental structure that AFLOW places below its reference is scored against the lowest structure of its own element but is not a corner of compound hulls (Supplementary Note 2). The models measure formation energies from their own lowest-energy structure of each element, so holding AFLOW's corners at the reference gives the AFLOW and model hulls the same corners. A structure is scored only in a system with at least three entries. The solver agrees with pymatgen's PhaseDiagram \cite{ong2013} to within 10\textsuperscript{--}\textsuperscript{1}\textsuperscript{5} eV on 80 sampled chemical spaces. The 25 meV/atom window is swept from 10 to 50 meV/atom (Supplementary Note 9). The Table 4 Alexandria row scores Alexandria itself as a ``model'' against the AFLOW hull, on the same-crystal pairs of the floor (Methods E). Alexandria covers only part of the benchmark, so we also score Alexandria and the models on the structures that AFLOW and Alexandria share as the same crystal (\texttt{scripts/\allowbreak alex\_\allowbreak ceiling\_\allowbreak shared.\allowbreak py}, rerun on these pairs by \texttt{scripts/\allowbreak dft\_\allowbreak ceilings\_\allowbreak sm.\allowbreak py} and \texttt{scripts/\allowbreak reffree\_\allowbreak common\_\allowbreak regen.\allowbreak py}).

\textit{Residual forces} are the maximum absolute force component that a model predicts at AFLOW's relaxed geometry, where the true forces are essentially zero (\texttt{fmax\_\allowbreak initial} in each relaxation record; \texttt{src/\allowbreak umlip\_\allowbreak bench/\allowbreak relax.\allowbreak py}).

\textit{Space-group retention and the symprec sweep} come from re-relaxing every benchmark structure with all fourteen models, with the production protocol on the production shards (\texttt{scripts/\allowbreak run\_\allowbreak shard\_\allowbreak symprec.\allowbreak py}). We report retention at symprec 0.01, 0.05, and 0.1 \AA{}, and assign the space group of AFLOW's structure and of the relaxed structure alike with spglib at the swept tolerance (\texttt{scripts/\allowbreak symprec\_\allowbreak reference\_\allowbreak sg.\allowbreak py}, \texttt{scripts/\allowbreak symprec\_\allowbreak full\_\allowbreak retention.\allowbreak py}). A structure enters a model's rate when its relaxation succeeded and both space groups exist; where retries left duplicate rows, the last one counts. Intervals are a percentile bootstrap over chemical systems. A looser tolerance can still give the two structures different higher-symmetry groups, so retention can dip slightly as the tolerance loosens; the largest dip is 0.4 points (MatterSim-1M). Retention is reported as a function of tolerance, never as a single number.

\textit{Polymorph energy ordering} is scored per composition. For each composition with at least three distinct structures whose AFLOW energies span at least 25 meV/atom, we compute the Spearman $\rho$ between the model's and AFLOW's formation energies, check whether the model's lowest-energy structure has AFLOW's minimum energy (ties count as hits), and report the median $\rho$ and the ground-state hit rate across compositions (\texttt{scripts/\allowbreak analyze\_\allowbreak polymorph\_\allowbreak ranking.\allowbreak py}).  The OQMD--Alexandria comparison (\texttt{scripts/\allowbreak polymorph\_\allowbreak floor\_\allowbreak oqmd\_\allowbreak alex.\allowbreak py}) pairs the two databases' structures by the same-crystal rule of Methods E, counts each same-crystal class once, and applies the same 25 meV/atom spread cut to OQMD's energies, on the 2,565 compositions with at least three shared polymorphs, grouped by reduced composition. Because 1,308 of those have exactly three, we report means, and compare them with AFLOW--Alexandria reweighted to the same set sizes. The AFLOW--OQMD comparison is built the same way, with OQMD's lowest-energy entry for each structure, on 2,318 compositions (\texttt{scripts/\allowbreak polymorph\_\allowbreak floor\_\allowbreak aflow\_\allowbreak oqmd.\allowbreak py}). OQMD's fitted elemental references shift every polymorph of a composition by the same constant and so leave its ordering unchanged. To compare models and databases like for like, we also score each model against AFLOW on exactly the structures a DFT pair uses, less any the model did not keep, re-apply the three-structure and 25 meV/atom rules, and score the second database on the same members. Intervals resample chemical systems.

\subsection*{H. Elastic response}
For the AEL subset, AFLOW stores the full DFT elastic tensor. Moduli are curvatures of the energy surface, so the comparison needs no energy reference. We recompute each compound's tensor with the model by a finite-strain protocol mirroring AEL's \cite{toher2017}: relax the structure, apply normal strains of $\pm$0.5\% and $\pm$1\% and shear strains of $\pm$1\% and $\pm$2\%, relax the ions at fixed cell for each deformation (FIRE, fmax 0.02 eV/\AA{}, at most 200 steps), and fit the tensor from the model's analytic stress in the ASE sign convention (no sign flip; a check showed that flipping the sign gives unphysical moduli; pymatgen \texttt{DeformedStructureSet} and \texttt{ElasticTensor.\allowbreak from\_\allowbreak independent\_\allowbreak strains}; \texttt{src/\allowbreak umlip\_\allowbreak bench/\allowbreak elastic.\allowbreak py}). From the tensor we read the Voigt--Reuss--Hill \cite{hill1952} bulk modulus K and shear modulus G and the Debye temperature \cite{toher2014}. We stratify by U status. Except in the LIB2 and LIB3 alloy libraries, AFLOW applies +U to every compound containing any of 28 elements (Sc--Ga, Nb--Sn and Ta--Au), whatever its anion (Supplementary Table 6 gives each population's +U fraction). The Materials Project's +U elements are eight of these, in oxides and fluorides only. Supplementary Note 12 splits the +U structures into oxides and fluorides with and without one of the Materials Project's eight elements, other anion compounds, and metallic compounds (\texttt{scripts/\allowbreak elastic\_\allowbreak u\_\allowbreak split.\allowbreak py}). Correlations are computed in log space (log\textsubscript{1}\textsubscript{0} predicted against log\textsubscript{1}\textsubscript{0} AFLOW; \texttt{scripts/\allowbreak elastic\_\allowbreak logspace.\allowbreak py}), which leaves out the structures a model predicts with a non-positive modulus (32--155 per model on the full set). We also report the log--log slope (ordinary least squares) and the median multiplicative bias (ten to the median of log\textsubscript{1}\textsubscript{0} predicted $-$ log\textsubscript{1}\textsubscript{0} AFLOW). The AEL ground truth lives mostly on ICSD compounds containing nonmetals, so this comparison is not metals-only. The metals-only recut is in Supplementary Note 12. The finite-strain task was run on the original eight-model panel, which is why it includes the two superseded checkpoints.

\subsection*{I. Bad references: detection and the third-database check}
\textit{The consensus test} works as follows. For each structure scored by at least six models of the eight-model panel (Methods A), we take every model's energies after the elemental-reference screen (Methods B) and compute the consensus C (the median model formation energy), the robust inter-model spread S (1.4826$\times$ the median absolute deviation), the distance dA = \textbar{}C $-$ $E_f(\mathrm{AFLOW})$\textbar{}, and the z-score dA / max(S, 0.02 eV/atom) (\texttt{scripts/\allowbreak secvi\_\allowbreak chain\_\allowbreak s\_\allowbreak dA\_\allowbreak z.\allowbreak py}). A model's formation energy on a structure is discarded as a numerical failure when it differs from AFLOW's by more than 50 eV/atom, and the 20 structures this leaves with fewer than six models are not scored. A structure on which AFLOW and the model consensus disagree beyond the cut (S < 0.1 eV/atom, dA > 0.3 eV/atom, and z $\ge$ 5; all strict inequalities on meV/atom values) is a \textit{discrepancy}. The test scores 1,419,025 structures. Sweeping z from 3 to 20 moves the count smoothly between 16,367 and 3,633, by about 5\% per unit around the setting (\texttt{scripts/\allowbreak flags\_\allowbreak vs\_\allowbreak threshold\_\allowbreak post.\allowbreak py}).

\textit{Re-referencing} fits one offset per element to the signed gap C $-$ $E_f(\mathrm{AFLOW})$, as a ridge regression ($\lambda$ = 10\textsuperscript{--}\textsuperscript{3}) on composition over the structures with \textbar{}C $-$ $E_f(\mathrm{AFLOW})$\textbar{} < 0.5 eV/atom (\texttt{scripts/\allowbreak reference\_\allowbreak cut\_\allowbreak aligned.\allowbreak jl}). We subtract the fitted offset and apply the same cut again (S < 0.1 eV/atom, dA > 0.3 eV/atom, z $\ge$ 5).  Two of the structures that remain were not discrepancies before re-referencing. The chain is \texttt{scripts/\allowbreak secvi\_\allowbreak chain\_\allowbreak s\_\allowbreak dA\_\allowbreak z.\allowbreak py} and \texttt{scripts/\allowbreak secvi\_\allowbreak chain\_\allowbreak structmatch.\allowbreak py}.

\textit{Third-database checks} adjudicate what remains. For each of the 10,861 discrepancies left after re-referencing, we take the lowest formation energy X among the entries of Alexandria, OQMD, and MP that are the same crystal as the discrepant AFLOW structure (Methods E). The database sides with the models if \textbar{}X $-$ C\textbar{} < \textbar{}X $-$ $E_f(\mathrm{AFLOW})$\textbar{}. As robustness checks we also count results from a prototype-fingerprint lookup with no structure check, and with a volume check (volumes per atom agreeing within 10\%) added to the structure check, and we rebuild the whole chain with a consensus from only the six MP-generation models, none trained on OMat24 (Supplementary Note 13; \texttt{scripts/\allowbreak secvi\_\allowbreak robustness\_\allowbreak smsg.\allowbreak py}).

\textit{The genuine residual} is the cases where Alexandria sides with AFLOW are genuine model errors. We count their enrichment in the 4d/5d refractory metals (Ir, Os, Tc, Re, Mo, W, Ru, Rh, Pt, Ta, Nb, Hf) against the cases where the models are right (\texttt{scripts/\allowbreak secvi\_\allowbreak chain\_\allowbreak structmatch.\allowbreak py}; per-element counts in \texttt{results/\allowbreak analysis/\allowbreak secvi\_\allowbreak post\_\allowbreak enrichment.\allowbreak csv}).

\textit{Corpus correlation} is measured on every structure scored by all ten models of the panel that predates eSEN and eqV2 (Table 1 less those two and the two superseded checkpoints), excluding any on which one of them misses by more than 5 eV/atom. On these structures we compare the mean Spearman $\rho$ of the models' per-structure errors over the 21 same-corpus pairs with that over the 24 cross-corpus pairs (Table 1 grouping), with an exact permutation test over all 210 corpus labelings, whose smallest attainable p is 1/210 (\texttt{scripts/\allowbreak model\_\allowbreak corr.\allowbreak jl}, \texttt{scripts/\allowbreak corpus\_\allowbreak contrast\_\allowbreak exact.\allowbreak py}).

\subsection*{J. Familiarity, novelty, and generalization}
We call a structure \textit{familiar} if it is present in a model's training corpus and \textit{novel} if it is not. A structure is present if it is the same crystal (Methods E) as a training structure, found by the same exhaustive comparison within each reduced formula. We tag every AFLOW structure against Materials Project relaxed structures, standing in for MPtrj, against the Alexandria PBE bulk set, and against the Alexandria parent structure of every OMat24 training row, since the rattled OMat24 frames themselves are triclinic (space group 1) almost without exception (\texttt{scripts/\allowbreak smfam\_\allowbreak match.\allowbreak py}, \texttt{scripts/\allowbreak smfam\_\allowbreak smsg.\allowbreak py}, \texttt{scripts/\allowbreak smfam\_\allowbreak tags.\allowbreak py}; \texttt{results/\allowbreak analysis/\allowbreak familiarity\_\allowbreak smsg.\allowbreak csv}, \texttt{familiarity\_\allowbreak fractions.\allowbreak csv}). We report familiar fractions over the 1,419,259 AFLOW structures that at least one model scores. Of these, 552,041 match an OMat24 parent, and all of them also match an Alexandria structure. A prototype fingerprint would have tagged 519,479 of the 712,394 structures that match Alexandria and called another 31,851 familiar that are not (\texttt{results/\allowbreak analysis/\allowbreak familiarity\_\allowbreak smsg\_\allowbreak summary.\allowbreak json}). For the novel-minus-familiar gap of Fig. 7, we tag each model against the public corpora that stand in for its training data (\texttt{scripts/\allowbreak familiarity\_\allowbreak per\_\allowbreak model.\allowbreak py}). The Materials Project stands in for MPtrj and for the proprietary MatterSim training set, and the full Alexandria set for the subsampled Alexandria used in fine-tuning, so a stand-in that is a superset of the training data can over-count familiarity. Pooled tagging would call every model familiar with the same structures and misstate the gaps (on the new-composition tier below, CHGNet's gap is 52 meV/atom pooled and 73 per model). A novel structure falls in one of three tiers: its composition is in training with a different structure; its composition is new but its chemical system is in training; or its chemical system is absent from training. Fig. 7 compares the familiar structures with all three tiers together and marks the second separately. Structures whose tier cannot be determined, and those without an AFLOW $E_{\mathrm{hull}}$, are left out of both groups. The per-model familiar fractions use \texttt{scripts/\allowbreak dedup\_\allowbreak omat24\_\allowbreak overlap.\allowbreak py}. The sensitivity of the familiar fractions to the matching tolerances is in \texttt{results/\allowbreak analysis/\allowbreak familiarity\_\allowbreak smsg\_\allowbreak tolerance.\allowbreak csv}.

\textit{Matched comparisons} contrast familiar and novel structures at matched properties by stratifying them and averaging the within-stratum difference in MAE, weighted by the pooled number of structures in each stratum (\texttt{scripts/\allowbreak matched\_\allowbreak ehull\_\allowbreak generalization.\allowbreak py}, \texttt{scripts/\allowbreak r1\_\allowbreak 2\_\allowbreak matched\_\allowbreak controls.\allowbreak py}). The strata are AFLOW $E_{\mathrm{hull}}$ bins (edges 25, 100, 400, and 1,600 meV/atom), AFLOW $E_f$ bins (edges $-$400, $-$200, $-$100, 0, 100, 200, 400, and 800 meV/atom), and each of these crossed with the number of elements. A stratum enters only when both groups hold at least 200 structures (100 for the five models trained without OMat24 or Alexandria, whose familiar sets hold about 33,200 structures), and we report an estimate only when the retained strata hold at least half of each group. To match on chemistry over the whole novel set, we reweight the familiar structures by entropy balancing \cite{hainmueller2012} so that the element frequencies, and optionally the $E_{\mathrm{hull}}$-bin and element-count distributions, equal those of the novel set exactly (\texttt{scripts/\allowbreak r1\_\allowbreak 2\_\allowbreak chemistry\_\allowbreak controls.\allowbreak py}). We also match exactly on chemical system crossed with $E_{\mathrm{hull}}$ bin (at least 5 structures per group), which 60--65\% of the OMat-trained models' novel structures permit and 2\% of the others'. Intervals resample chemical systems (Methods D), with the balancing weights re-solved in each replicate.

\textit{Near the hull and across databases}, within 10 meV/atom of the hull, we compare each model's per-structure \textbar{}error\textbar{} distribution to the AFLOW--Alexandria disagreement on the same window and bins over the 25,546 of the 29,175 near-hull common-set structures (Methods G) whose Alexandria partner is the same crystal, chosen without reference to any model's results (\texttt{scripts/\allowbreak near\_\allowbreak hull\_\allowbreak yardstick\_\allowbreak common.\allowbreak py}). The cross-database test runs all fourteen models on 34,681 metallic compounds that were in OQMD but not in AFLOW when the set was built (180 of them match a later AFLOW fingerprint table). It scores the models against OQMD (\texttt{scripts/\allowbreak analyze\_\allowbreak oqmd\_\allowbreak aligned\_\allowbreak trim.\allowbreak py}) on the 12,409 of 34,681 whose elements all carry unary references (Methods B; 12,408 scored for SevenNet-0), with per-element re-alignment fit in-sample on the scored set by least absolute deviation, as in Methods F but without the held-out split, and the same trim. We split the scored compounds by whether they are the same crystal as a structure of any training corpus above (\texttt{scripts/\allowbreak oqmd\_\allowbreak generalization\_\allowbreak split.\allowbreak py}; \texttt{results/\allowbreak analysis/\allowbreak oqmd\_\allowbreak familiarity\_\allowbreak smsg.\allowbreak csv}): 7,588 match at least one and 4,821 match none. To bound memorization, a difference-in-differences test compares each model's improvement on a group of seen compounds, relative to the never-seen ones, between models trained on that group and models that were not (\texttt{scripts/\allowbreak oqmd\_\allowbreak memorization\_\allowbreak did.\allowbreak py}). For the compounds that only Alexandria contains, the four models trained on Alexandria are compared with the four OMat24-only models; for the Materials Project compounds, the five models trained only on the Materials Project are compared with the same four. The never-seen compounds are reweighted to the seen group's composition, by exact strata of element count and crystal system or by a logistic propensity, and the bootstrap resamples chemical systems jointly across models.

\textit{OMat24 held-out forces} are scored for all fourteen models on 50,002 configurations of OMat24's held-out validation split, a subsample drawn by uniform stride within each sub-dataset so that rattled and molecular-dynamics frames keep their natural proportion (\texttt{scripts/\allowbreak omat24\_\allowbreak subsample.\allowbreak py}). Each configuration is a single-point evaluation with no relaxation (\texttt{scripts/\allowbreak score\_\allowbreak omat24.\allowbreak py}). We report the per-component force MAE in eV/\AA{}, averaged over configurations after setting aside those whose force MAE exceeds 1 eV/\AA{}, overall and binned by reference force and frame type (\texttt{scripts/\allowbreak omat24\_\allowbreak aggregate.\allowbreak py}; \texttt{results/\allowbreak analysis/\allowbreak omat24\_\allowbreak force\_\allowbreak mae.\allowbreak csv}, \texttt{omat24\_\allowbreak force\_\allowbreak binned.\allowbreak csv}).

\subsection*{K. Compute cost}
Production ran the large models on GPU and the small ones on CPU. For comparability, the cost figure (Fig. 11; Supplementary Note 17) uses CPU-measured relaxation time throughout---the median per-structure cost in core-seconds (wall-clock seconds $\times$ allocated cores: 8 for the production array including UMA, 12 for GRACE-2L; \texttt{scripts/\allowbreak timing\_\allowbreak extract.\allowbreak py}), with eSEN re-measured on a matched CPU sample (\texttt{scripts/\allowbreak cpu\_\allowbreak timing\_\allowbreak sample.\allowbreak py}). We omit eqV2: its direct-force head barely relaxes (Methods C), so its relaxation time is not comparable. The full study used about 372,000 CPU core-hours and 830 GPU-hours on Brigham Young University's research cluster, roughly half for the production relaxations and half for re-relaxing the whole benchmark to measure space-group retention.

\subsection*{L. Use of AI assistance}
The analysis code, figure scripts, and manuscript drafts were developed with the assistance of an AI coding agent (Claude Code, Anthropic; Claude Opus 4.8, Claude Sonnet 4.6, Claude Fable 5, Claude Opus 5, and Claude Opus 5.5, June--October 2026) operating under the authors' direction. The authors specified every analysis and made every scientific decision, but did not review the generated code line by line. Correctness was established instead by checks that do not depend on reading it. The manuscript's numbers were given written specifications taken from the text, and separate agents, working from the raw data without seeing the printed values, wrote new code to recompute them. 801 numbers were re-derived this way, and every mismatch was traced through the original scripts and resolved. Afterward, a regenerated result replaced an earlier one only after the new pipeline had reproduced the earlier output exactly, and a separate AI review, instructed to act as an additional referee, compared the manuscript's claims with the data files. The exhaustive structure matching of Methods E and the analyses regenerated from it came after these audits and passed the same kind of checks: each script reproduced its earlier output before switching to the new pairs, and an independent re-derivation confirmed the new numbers. The authors reviewed every change to the text. Each reported number is produced by a released script and traces to the released data, and environment lockfiles together with a checkpoint-provenance manifest pin the software stack. The authors take full responsibility for the work.

\clearpage\onecolumn
\includepdf[pages=-]{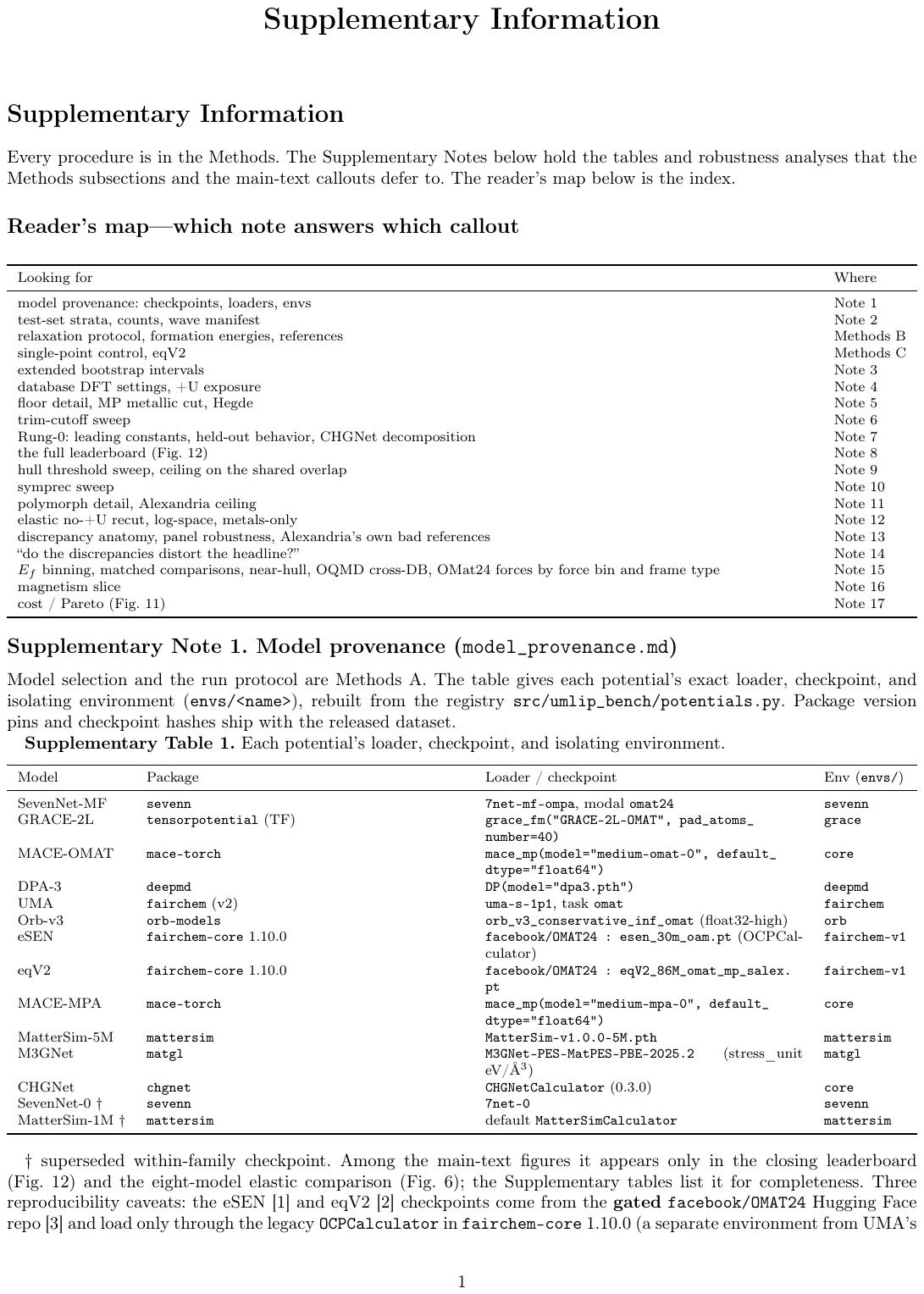}
\end{document}